\documentclass[final,3p,times,11pt]{elsarticle}
\usepackage{graphicx}
\usepackage{array}
\usepackage{enumitem}
\usepackage[utf8]{inputenc}
\usepackage{adjustbox}
\usepackage{booktabs}
\usepackage{siunitx}
\usepackage{float}
\usepackage{setspace}

\usepackage[normalem]{ulem} 
\usepackage{xcolor}

\usepackage{eurosym}
\setlist[itemize]{leftmargin=0pt,labelsep=0.6em,itemsep=0.2em,topsep=0.3em,parsep=0pt}
\setlist[enumerate]{leftmargin=0pt,labelsep=0.6em,itemsep=0.2em,topsep=0.3em,parsep=0pt}
\newlist{flatroman}{enumerate}{1}
\setlist[flatroman]{label=\textit{\roman*)},leftmargin=0pt,labelsep=0.6em,align=parleft}

\usepackage[table]{xcolor}
\usepackage{multirow}
\usepackage{amssymb}
\usepackage{subcaption} 
\usepackage{algorithm}
\usepackage{amsthm}
\usepackage{algpseudocode}
\usepackage{threeparttable} 
\usepackage{pgfplots}
\usepackage{pgfplotstable}
\usepackage{tikz}
\pgfplotsset{compat=1.18}
\usepgfplotslibrary{groupplots}

\definecolor{summercol}{RGB}{76,153,0} 
\definecolor{wintercol}{RGB}{230,115,0} 
\usepackage{makecell}
\usepackage{empheq}
\definecolor{lightgreen}{RGB}{220, 255, 220}
\definecolor{lightorange}{RGB}{255, 240, 220}
\biboptions{sort&compress}
\usepackage{appendix}
\usepackage{bbm}
\usepackage{tabularx}
\usepackage{url}
\usepackage{float}
\usepackage{textcomp}
\usepackage[T1]{fontenc}
\usepackage{lmodern}
\usepackage{pdflscape}
\newtheorem{definition}{Definition}
\usetikzlibrary{arrows.meta,positioning,fit,backgrounds,shapes,shapes.multipart,shadows.blur,matrix}

\usepackage{titlesec}
\titlespacing{\section}{0pt}{1.0ex plus.2ex minus.2ex}{0.5ex}
\titlespacing{\subsection}{0pt}{0.8ex plus.2ex minus.2ex}{0.4ex}
\tikzset{arrowstyle/.style={draw=black, thick, -Latex}}
\usepackage{orcidlink} 
\usepackage{nomencl}
\makenomenclature

\usepackage{mathtools}

\newdefinition{rmk}{Remark}
\newtheorem{ass}{Assumption}
\newtheorem{prop}{Proposition}
\newtheorem{prof}{Proof}
\usepackage{hyperref}
\newcommand{\sref}[2]{\hyperref[#2]{#1 \ref*{#2}}}

\let\oldref\ref
\renewcommand{\ref}[1]{(\oldref{#1})}
\usepackage[nameinlink, capitalise, noabbrev]{cleveref}
\hypersetup{
colorlinks,
citecolor=green,
filecolor=green,
linkcolor=green,
urlcolor=green}
\crefname{prop}{Proposition}{Propositions}
\crefname{thm}{Theorem}{Theorems}
\crefname{cor}{Corollary}{Corollaries}
\crefname{ass}{Assumption}{Assumptions}

\journal{OMEGA}

\allowdisplaybreaks
\begin{document}
\begin{frontmatter}

\title{Bidding strategies for energy storage operators in a 100\% renewable electricity market: A game-theoretic approach}

\author{Arega Getaneh Abate\orcidlink{0000-0002-5517-2585}\corref{cor1}}
\ead{ageab@dtu.dk}
\author{Dogan Keles\orcidlink{0000-0002-9620-6294}}
\ead{dogke@dtu.dk}
\author{Xiufeng Liu\orcidlink{0000-0001-5133-6688}}
\ead{xiuli@dtu.dk}
\author{Salim Hassi}
\ead{salim@gmail.com}
\author{Xiao-Bing Zhang\orcidlink{0000-0001-8155-5769}}
\ead{xiazhan@dtu.dk}
\cortext[cor1]{Corresponding author}
\address{Department of Technology, Management and Economics, Technical University of Denmark, Kgs. Lyngby, Denmark}

\begin{abstract}
\noindent
  Large-scale energy storage is expected to be a pivotal source of flexibility in electricity systems supplied entirely by renewable energy sources (RES). However, its strategic role in market-based dispatch remains insufficiently understood. In a 100\% RES market setting, storage can improve adequacy and renewable utilization by shifting energy across time. It can also acquire market power as the main flexible, price-making technology. In this paper, we develop a Cournot competition model in which storage operators choose quantity bids to maximize profit in a stylized day-ahead electricity market supplied only by RES. Market clearing is represented through residual-demand blocks so that renewable intermittency appears as an intertemporal arbitrage opportunity for storage operators. We formulate the storage operators' problem in two tractable ways: \emph{i)} a continuous reformulation based on demand blocks, and \emph{ii)} an equivalent mixed-integer linear programming (MILP) model with big-$M$ linearization. Nash equilibria are computed with an iterative best-response procedure and benchmarked against a centralized social planner problem to quantify efficiency losses from strategic behavior. The model is calibrated to Denmark's DK1 bidding zone using 2024 day-ahead data and 2030 renewable-capacity and demand projections. The results show that storage reduces imbalances and improves welfare relative to a no-storage case. Concentrated ownership also creates incentives to withhold flexibility, raise prices, and slow the reduction of unmet demand and curtailment. These findings position storage as both a stabilizing resource and a potential source of market power, highlighting the importance of market designs that jointly consider competition, concentration, and capacity deployment in high-RES systems.
\end{abstract}

\end{frontmatter}
\vspace{1mm}

\section{Introduction}

\vspace{1mm} 

The transition toward an electricity system dominated by renewable energy sources (RES) is accelerating, driven by decarbonization, energy-security concerns, and rapid cost reductions in wind and solar technologies \citep{zhong2021towards}. However, high RES shares also make supply-demand balancing more difficult because RES output is intermittent, weather-dependent, and only imperfectly aligned with demand peaks \citep{vivas2018review}. As the system approaches a 100\% renewable configuration, these temporal mismatches transform from a peripheral balancing issue into a central market-design and operations problem \citep{liu2022novel}.
To that end, energy storage is one of the most important technologies for addressing this mismatch. By charging during hours of surplus generation and discharging during hours of scarcity, storage can reduce curtailment, limit unmet demand, smooth prices, and improve the market value of RES generation \citep{johnson2024optimal,yildiran2023robust}. Nevertheless, as storage capacity expands, storage owners do not remain passive balancing resources. They become strategic market participants whose bidding and scheduling decisions influence prices, welfare, and the effective value of flexibility \cite{grimm2021optimal,hao2020does}.

This strategic dimension is especially relevant in concentrated storage markets. When a small number of operators control a large share of flexible capacity, they may find it profitable to withhold discharge, delay charging, or otherwise exploit scarcity in ways that increase profits but reduce social welfare. Existing studies address related effects in a range of electricity-market settings. For example, \cite{gonzalez2021transmission} study competition in a generation-transmission expansion framework, while \cite{bjorndal2023energy} examine how a strategic storage operator affects market outcomes under alternative market designs. Other studies assess strategic storage operation, sizing, and arbitrage under different assumptions about price formation, uncertainty, and network structure \citep{zhao2022strategic,Huang2022,zhang2023optimal}. In a planning-oriented complement, \cite{misicc2025optimal} optimize the sizing and siting of storage in transmission grids connected to wind farms. Nevertheless, the literature still offers limited game-theoretic interactions on \emph{multiple} competing storage operators in a purely RES-supplied electricity market in which renewable generators bid their available output at or near zero marginal cost, leaving storage as the main flexible, price-making resource.

This gap matters from both theoretical and market design perspectives. In such settings, the relevant questions are not only whether storage improves system performance, but also how ownership concentration, capacity allocation, and operating assumptions shape equilibrium prices, dispatch, and welfare. We address this gap by developing a Cournot competition model for a 100\% renewable electricity market, organized around three specific research questions (RQs):
1) \textit{How does competition among storage operators affect prices, dispatch patterns, and social welfare?}
2) \textit{To what extent does market power among storage operators induce market inefficiencies?}
3) \textit{Under what conditions do privately optimal storage operations align with social welfare maximization?} Our objective is not to replicate all technical details of a full market-clearing algorithm. Instead, we deliberately construct a stylized model that isolates the strategic role of storage in a day-ahead market with a 100\% RES supply. We therefore model storage operators as quantity-setting players in a Cournot game. In this context, price-based competition is less informative because renewable generators bid at or near zero marginal cost, whereas storage exerts market power primarily through intertemporal quantity shifts. To ensure the reliability of these insights, we systematically evaluate how core modeling choices, such as demand-curve construction and operational boundaries, affect market outcomes. 

As quantity-setting players, storage operators choose profit-maximizing charge and discharge schedules while anticipating their rivals' actions. A centralized social planner benchmark provides a welfare reference and quantifies the efficiency losses from strategic behavior. The proposed model is calibrated to Denmark's DK1 bidding zone using 2024 RES profiles, Danish Energy Agency (DEA) 2030 capacity projections, and realistic parameters. Unlike settings with monopolistic storage operators or co-existing conventional flexibility \citep{gaudard2019energy,khalid2016minimizing}, the market considered in this model is supplied entirely by RES. Although our market setup is stylized, it provides a useful benchmark for future market-design discussions. In particular, in a market 100\% supplied by renewable energy sources, storage becomes indispensable for supply-demand balancing and therefore represents a plausible source of strategic market power. This stylized setting nonetheless remains rich enough to implicate practical insights into ownership structure, storage sizing, and market design. Thus, the main contributions of this study are fourfold:
\begin{enumerate}
\item[\textit{i)}] To address RQ1, we develop a Cournot competition model for storage in a day-ahead electricity market under a 100\% RES system. Intermittency is captured through residual demand and interpreted as an intertemporal arbitrage opportunity. Each storage operator chooses charge and discharge schedules while internalizing the effect of its quantities on the market-clearing price. This allows us to identify how storage competition shapes dispatch, scarcity, and price formation in a renewables-dominated system.

\item[\textit{ii)}] To address RQ2, we formulate a centralized social planner benchmark and two tractable optimization formulations for the strategic problem: \emph{i)} a continuous reformulation based on demand blocks, and \emph{ii)} a discrete MILP with big-$M$ constraints. This modeling structure makes it possible to precisely quantify welfare losses from strategic behavior while also clarifying the computational implications of different linearization choices. 

\item[\textit{iii)}] To address RQ3, we evaluate the robustness of these market dynamics under varying structural assumptions. By systematically analyzing the directional bias of our modeling choices, such as the representative-day selection, and terminal state of charge (SoC), we establish the boundary conditions of our results. This ensures that the core mechanisms of capacity withholding and diminishing welfare returns are disentangled from specific data-processing artifacts.

\item[\textit{iv)}] We develop a data-driven case study for Denmark's DK1 bidding zone to derive managerial and policy insights. In particular, the analysis maps market concentration and aggregate storage capacity to welfare, price behavior, and scarcity, thereby clarifying when deploying additional storage capacity is socially valuable and when ownership structure becomes the more important regulatory instrument.
\end{enumerate}

The remainder of the study is structured as follows. Section \ref{lit} reviews the literature on storage bidding strategies. Section \ref{method} presents the mathematical model and the solution approaches. Section \ref{design} describes the Denmark case study and the simulation results. Section \ref{sec_discussion} discusses key insights, robustness of the results, and outlines some limitations. Section \ref{conclusion} concludes the paper. 

\section{Literature review}\label{lit}

Bidding strategies for storage facilities in electricity markets have been studied extensively across a wide range of system contexts and modeling frameworks \cite{xie2025strategic,mei2024two}. 
An important literature examines storage as an intertemporal resource for commitment, scheduling, and system operation under intermittent supply and renewable uncertainty, including optimal energy commitments with storage \citep{kim2011optimal}, unit commitment under correlated renewable uncertainty \citep{cordera2023unit}, and general storage models for energy systems \citep{lust2019general}. In the literature, two broad bidding approaches emerge: \textit{economic bidding}, which submits both price and quantity bids, and \textit{self-scheduling}, where the storage operator commits only to specific energy quantities, irrespective of market-clearing prices \cite{finnah2022integrated}.

Economic bidding problems are typically formulated as bi-level optimization problems, where the upper level represents the storage operator's profit-maximizing behavior and the lower level models the market-clearing problem solved by the market operator \cite{nasrolahpour2016bidding}. This bilevel structure has been used to evaluate grid-scale storage arbitrage under wind generation and locational-price smoothing \cite{cui2017bilevel}, and to coordinate the price-maker operation of large storage units in nodal energy markets \cite{8Coordinated}. More recent bilevel formulations extend it to a storage agent bidding jointly into energy and reserve markets under stochastic generation \cite{dimitriadis2022strategic}, and to the joint investment and bidding of wind producers across electricity and green-certificate markets \cite{kanta2025optimal}. Economic bidding is particularly suitable in deterministic environments, where arbitrage opportunities arise predictably across adjacent time intervals.

Self-scheduling studies can be broadly grouped into three categories according to how they model the interaction between storage operations and market prices. The \textit{first method} uses historical prices and assumes storage operators act as pure price-takers with no influence on market outcomes (e.g., \cite{zhang2026optimal,keles2022evaluation}). The \textit{second method} models price as a function of total energy supplied to the market, typically derived from historical data. 
A common assumption in this line of work is a linear price-quantity relationship \cite{10Cournot,zhang2020cournot}. The \textit{third method}, which is also the most prevalent, assumes inelastic electricity demand and thereby assigns market power to generators. In this setting, the aggregated supply curve determines the market price. This assumption is reasonable in contexts where day-to-day demand fluctuations are relatively insensitive to price. 
For example, \cite{going} develop a self-scheduling model for a storage aggregator in the Great Britain market and show that more intensive grid-service participation lowers household electricity bills. 
 Along similar lines, \cite{barbry2019robust} construct upper and lower residual supply curves to formulate a robust optimization problem for storage self-scheduling, incorporating nodal constraints in the New York electricity market.

From a methodological perspective, an increasing number of studies adopt game-theoretic formulations to capture competition among multiple storage entities. 
In \cite{zhao2022strategic}, players compete first on investment decisions related to power capacity and storage size, and then on arbitrage revenue generation over a multi-year horizon in the Californian market. They demonstrate that the relative efficiency of storage operators plays a critical role in shaping their market power. Similar findings are reported in \cite{10Cournot}, which models a game between multiple storage operators, who maximize their individual profits, and a market operator aiming to maximize social welfare in a multi-node setting. 
 The authors show that although the inclusion of storage improves social welfare, competitive bidding prevents the system from reaching the socially optimal outcome \cite{nasrolahpour2016bidding}. A Cournot oligopoly game involving multiple generators, including storage players, competing to meet electricity demand, is modeled in \cite{zhang2020cournot}. That study formulates a Cournot oligopoly model with storage, wind, PV, and thermal units that incorporates uncertainty penalty costs and computes a unique Nash equilibrium using a Newton-based optimal generation plan algorithm. 
\cite{motalleb2017non} propose a game-theoretic market framework in which demand-response aggregators compete to sell energy from consumer storage, deriving optimal bidding strategies under incomplete information and varying market conditions. Relative to the literature discussed above, this game-theoretic stream shifts attention from optimal storage use under uncertainty to strategic behavior, endogenous price formation, and welfare under decentralized ownership.

The scheduling horizon in these models ranges from a single hour to multiple years. Therefore, short-term models are common operational decision tools. For instance, a 1-hour horizon is used in the Cournot local-energy-trading model of \cite{zhang2020cournot}, whereas a 2-day horizon is adopted in the storage-aggregator model of \cite{going}. The most common choice is a 1-day horizon, which appears in single-facility storage bidding \cite{nasrolahpour2016bidding}, robust self-scheduling in the New York market \cite{barbry2019robust}, networked-Cournot storage operation \cite{10Cournot}, pumped-hydro market-power analysis \cite{pump}, and imbalance-option trading with storage \cite{szabo2020optimal}. These short-term models typically assume perfect foresight of prices and production, with the exception of \cite{barbry2019robust}. Forecasts for renewable-energy generation and electricity demand are generally reliable over such short-term intervals \citep{forecast}. However, boundary effects may limit operational flexibility and arbitrage revenues \citep{cui2017bilevel,pump}. In contrast, longer-term models spanning a week \citep{cui2017bilevel}, a year \citep{nasrolahpour2016bidding}, or multiple years \citep{zhao2022strategic} may more accurately capture realistic bidding strategies and storage behavior over time. However, their relevance depends on the market design and auction mechanisms in which storage operators participate. 

While several studies have examined scenarios with significant renewable penetration \citep{cui2017bilevel,zhao2022strategic}, and the broader literature has established the importance of storage for commitment and system operation under intermittency and uncertainty \citep{kim2011optimal,cordera2023unit,lust2019general}, few have thoroughly explored market dynamics in a completely renewable-powered electricity system where storage is the sole flexible and price-making resource. For instance, \cite{Huang2022} model storage operators under Cournot competition, but their framework ignores the typical European zonal bidding system and does not account for the 100\% RES supply dilemma of price dampening and energy insecurity.

This paper addresses that gap by developing a self-scheduling Cournot model for a 100\%-RES day-ahead market in which renewables bid (near) zero at their available capacity and storage is the only strategic, price-making technology. We assume perfect foresight of hourly demand curves and renewable production and let multiple, heterogeneous storage operators submit quantity-only bids (charge/discharge schedules) while internalizing their impact on prices. We provide two tractable formulations: \textit{i}) a big-$M$ MILP and \textit{ii}) a continuous-variable reformulation. Market outcomes are benchmarked against a centralized social-planner problem to quantify efficiency losses from strategic behavior. The proposed model combines endogenous pricing with intertemporal storage constraints to capture arbitrage, adequacy, and curtailment effects, yielding policy-relevant insights into how market design and storage sizing or participation can improve welfare in future renewables-dominated systems. Table~\ref{tab_lit} positions the proposed model against the existing literature.

\begin{table}[htbp]

\centering
\caption{Positioning of the proposed model relative to representative strategic-storage studies. \checkmark~indicates the feature is present, and $\times$~indicates that it is absent or not modeled.}
\label{tab_lit}
\footnotesize
\begin{adjustbox}{width=\textwidth}
\begin{tabular}{@{}l l c c c c c l@{}}
\toprule
\textbf{Study} & \textbf{Framework} & \textbf{\makecell{Multiple\\operators}} & \textbf{\makecell{Price-maker\\storage}} & \textbf{\makecell{100\% RES\\supply}} & \textbf{\makecell{Endogenous\\demand curve}} & \textbf{\makecell{Planner\\benchmark}} & \textbf{Solution method} \\
\midrule
\cite{nasrolahpour2016bidding}   & Bilevel & $\times$ & \checkmark & $\times$ & $\times$ & $\times$ & MPEC / MILP \\
\cite{cui2017bilevel}            & Bilevel & $\times$ & \checkmark & $\times$ & $\times$ & $\times$ & Bilevel LP \\
\cite{zhao2022strategic}         & Cournot (investment) & \checkmark & \checkmark & $\times$ & $\times$ & $\times$ & Equilibrium search \\
\cite{10Cournot}                 & Networked Cournot & \checkmark & \checkmark & $\times$ & \checkmark & \checkmark & Networked Cournot \\
\cite{zhang2020cournot}          & Cournot oligopoly & \checkmark & \checkmark & $\times$ & $\times$ & $\times$ & Newton-based NE \\
\cite{bjorndal2023energy}        & Bilevel (monopoly) & $\times$ & \checkmark & $\times$ & \checkmark & \checkmark & Bilevel / MILP \\
\cite{dimitriadis2022strategic}  & Bilevel (stochastic) & $\times$ & \checkmark & $\times$ & $\times$ & $\times$ & Stochastic MPEC \\
\cite{kanta2025optimal}          & Bilevel (investment) & $\times$ & \checkmark & $\times$ & $\times$ & $\times$ & Bilevel MILP \\
\cite{anunrojwong2024battery}    & Strategic / Cournot & \checkmark & \checkmark & \checkmark & \checkmark & \checkmark & Analytical (price of anarchy) \\
\textbf{This paper}              & Cournot (self-scheduling) & \checkmark & \checkmark & \checkmark & \checkmark & \checkmark & MILP + continuous reformulation \\
\bottomrule
\end{tabular}
\end{adjustbox}

\end{table}

\section{Problem description, formulations, and solution methods} \label{method}

This section describes the detailed mathematical formulation of the proposed storage self-scheduling Cournot model. 
The problem formulation is organized into two parts: \textit{Cournot competition model} and \textit{Social planner problem formulation}. The system-level problem is considered as a benchmark, where the social planner optimizes a coordinated welfare-maximization problem while disregarding competition among storage players.

\subsection{Problem description and assumptions}\label{subsec_assumptions}

Our goal is to study strategic interactions among storage operators in a \emph{stylized day-ahead} electricity market supplied by a 100\% RES with near-zero marginal costs. In the absence of flexible conventional supply, storage becomes the main technology capable of shifting energy intertemporally and moderating renewable intermittency. We therefore represent market clearing through a residual-demand curve described by price-quantity blocks and interpret storage operation as arbitrage between renewable surplus and renewable scarcity. The resulting equilibrium is compared with a centralized social planner solution to evaluate efficiency losses from strategic behavior. 

\begin{ass}\label{ass_behavior}
Storage operators are modeled as risk-neutral profit maximizers who play a non-cooperative, simultaneous-move quantity game (Cournot competition) in the day-ahead market, taking their rivals' hourly charge and discharge decisions as exogenous. Furthermore, we assume storage operators possess perfect foresight regarding the residual-demand blocks and hourly RES availability used in clearing, treating RES intermittency as a deterministic intertemporal arbitrage opportunity.
\end{ass}

\cref{ass_behavior} formalizes storage as a quantity-setting technology. With a 100\% renewable supply, price- or cost-based competition is less informative because renewable offers are concentrated near zero marginal cost. In an oligopolistic setting, strategic behavior arises primarily through withholding or shifting quantities across hours. We intentionally employ this assumption to cleanly isolate the strategic mechanisms under study from the confounding effects of forecast errors. Conflating these two mechanisms in our baseline model would preclude us from accurately attributing the observed market outcomes to strategic interactions rather than to information asymmetry. While an abstraction, perfect foresight serves as the standard informational baseline in the strategic-storage and Cournot-equilibrium literature. Early studies adopt it for pumped-hydro market-power analysis \cite{Schill2010} and single-facility storage bidding \cite{nasrolahpour2016bidding}. More recent work retains it when analyzing imperfect competition and the price of anarchy in battery operation \cite{Huang2022,anunrojwong2024battery}.

\begin{ass}\label{ass_scope}
The market is modeled as a single bidding zone, and each player $p$ decides without transmission constraints. The analysis focuses on the day-ahead energy market, in which energy is cleared through aggregate price-quantity blocks under uniform pricing. Balancing reserves, ancillary-service activation, and capacity remuneration are not part of this clearing problem and are therefore outside the model scope.
\end{ass}

\cref{ass_scope} isolates strategic storage interaction in a stylized day-ahead energy-market environment. This abstraction does not aim to capture congestion-induced locational market power or cross-market interactions between day-ahead trading and real-time system balancing. Accordingly, the quantitative magnitudes may differ in nodal, multi-zone, or multi-product market settings. Under this market design, RES generators offer all available energy at (near-)zero marginal cost, and the number and size of storage operators are treated as exogenous because the focus is on short-run operation.

These assumptions are analytically useful because they provide a clean benchmark for identifying the strategic role of storage in a renewable-dominated day-ahead market. By abstracting from network constraints, the model focuses on market-wide price formation rather than congestion-driven locational scarcity or node-specific storage value. Perfect foresight improves the timing of charging and discharging decisions, but it also allows the analysis to separate strategic interaction from forecasting error and thus makes the underlying competitive mechanism more transparent. The results should therefore be interpreted as stylized market outcomes, with forecast errors and network constraints affecting the quantitative magnitudes rather than the basic economic mechanism. 

\begin{table}
\footnotesize
\caption{Notation used throughout the paper.}
\begin{tabular}{@{}ll@{}}
\toprule
\textbf{Notation} & \textbf{Description} \\ \midrule
\textbf{Indices} & \\
$t$ & Index of time periods running from $1$ to $T$ \\
$i$ & Index of power levels running from $1$ to $N$ \\
$j$ & Index of demand-curve bid blocks running from $1$ to $D$ \\
$p$ & Index of players, running from $1$ to $P$ \\
$o$ & Index of rival players, with $o\neq p$ \\
$k,m$ & Auxiliary indices for demand-curve blocks in summations \\
\textbf{Parameters} & \\
$OC_p$ & Operating cost for storage player $p$ \\
$Q_p^{\max}$ & Maximum charging/discharging power for player $p$ \\
$E_p^{\max}$ & Maximum battery capacity for player $p$ \\
$q_{tj}^{\max}$ & Power volume of demand block $j$ at hour $t$ \\
$Q_{i}^{p}$ & Power volume at level $i$ for player $p$ \\
$q_{tj}^{\min}$ & Cumulative power volume from demand blocks $1$ to $j-1$ at hour $t$\\
$q_t^{\mathrm{ch},o}$ & Aggregate charging quantity of rival players at time $t$ \\
$q_t^{\mathrm{dis},o}$ & Aggregate discharging quantity of rival players at time $t$ \\
$\eta^p$ & Efficiency parameter of player $p$ (applied to charging) \\
$\alpha^{\mathrm{batt}}$ & Initial state-of-charge (SoC) fraction of the battery \\
$\epsilon$ & Tolerance ratio for final battery level \\
$\mathrm{RES}_t$ & Renewable energy production at time $t$ \\
$\mathrm{pr}_{tj}$ & Price of demand block $j$ at time $t$ \\
$\mathrm{vol}_{tj}$ & Cumulative demand volume up to block $j$ at time $t$ \\
$V_t^{\max}$ & Maximum cumulative demand volume at hour $t$ \\
$M_t$ & Time-specific big-$M$ constant used in the MILP formulation \\
$\Delta \mathrm{vol}_{tk}$ & Incremental demand volume of block $k$ at time $t$ \\
\textbf{Variables} & \\
$e^{p}_t$ & Battery energy level of player $p$ at time $t$ \\
$z^{\mathrm{ch},p}_{ti}$ & Binary: charging decision $i$ of player $p$ at time $t$ \\
$z^{\mathrm{dis},p}_{ti}$ & Binary: discharging decision $i$ of player $p$ at time $t$ \\
$b_{tj}$ & Continuous quantity served within demand block $j$ at time $t$ \\
$u_{tj}$, $u^{+}_{tj}$, $u^{-}_{tj}$ & Binary variables for the active demand block and the above/below-range indicators \\
$w^{\mathrm{ch},p}_{tij}$, $w^{\mathrm{dis},p}_{tij}$ & Auxiliary binary variables linking the active demand block and the charging/discharging action \\
$q^{\mathrm{ch},p}_{t}$, $q^{\mathrm{dis},p}_{t}$ & Charge/discharge quantity of player $p$ at time $t$ \\
$\lambda_{t}$ & Market price at time $t$ \\
$y_{tj}$ & Binary indicator equal to $1$ if demand block $j$ lies above the market-clearing price at time $t$ \\
\textbf{Derived quantities} & \\
$q_t^{\mathrm{tot}}$ & Net quantity supplied to the day-ahead market at time $t$ \\
$q_t^{*}$ & Market-cleared quantity at time $t$ \\
$u_t$ & Unmet demand (scarcity) at time $t$ \\
$cur_t$ & Curtailed renewable production at time $t$ \\
$\Pi_t^p$ & Profit of storage player $p$ at time $t$ \\
$\Pi_t^{\mathrm{RES}}$ & Renewable producer surplus at time $t$ \\
$\mathrm{CS}_t$ & Consumer surplus at time $t$ \\
$\mathrm{PS}_t$ & Producer surplus at time $t$ \\
\textbf{Decision-variable sets} & \\
$\chi^p$ & Set of decision variables in the big-$M$ formulation for player $p$ \\
$\chi_{\mathrm{alt}}^p$ & Set of decision variables in the alternative formulation for player $p$ \\
$\chi^{\mathrm{SO}}$ & Set of decision variables in the social-planner formulation \\
\bottomrule
\end{tabular}
\label{TAB_1}
\end{table}

\subsection{Cournot competition model}

Given Assumptions \ref{ass_behavior} and \ref{ass_scope}, this subsection formulates each storage operator’s best-response problem under Cournot competition. Each operator chooses an hourly charging and discharging schedule to maximize profit subject to battery constraints and endogenous market clearing. To keep the strategy space finite and computationally tractable, charging and discharging
actions are discretized into $N$ positive levels plus the idle action. The levels are equally spaced multiples of $Q_p^{\max}/N$, where $Q_p^{\max}$ is the maximum charging or discharging rate of player $p$. Binary variables $z^{\mathrm{ch},p}_{ti}$ and $z^{\mathrm{dis},p}_{ti}$ indicate the selected charging and discharging decisions at hour $t$. The residual-demand curve is likewise approximated by $D$ stepwise bid blocks, and the binary variable $u_{tj}$ identifies the active demand block at hour $t$. Table \ref{TAB_1} summarizes the main notations used in the paper. 

\begin{definition}\label{def_levels}
Given the discrete power levels (equally spaced levels up to the technical maximum $Q_p^{\max}$) $Q_i^p:= \frac{Q_p^{\max}}{N}\,i$ for $i=1,\ldots,N$, the total charge and discharge of player $p$ at time $t$ are

$$q^{\mathrm{ch},p}_t=\sum_{i=1}^N Q_i^p z^{\mathrm{ch},p}_{ti},\qquad
q^{\mathrm{dis},p}_t=\sum_{i=1}^N Q_i^p z^{\mathrm{dis},p}_{ti},$$

where $z_{ti}^{\mathrm{ch},p}$ and $z_{ti}^{\mathrm{dis},p}$ are binary variables representing charging and discharging decisions of player $p$ at hour $t$. Note that $q_{t}^{\mathrm{ch},p}$ and $q_{t}^{\mathrm{dis},p}$ are the decision variables of player $p$ at hour $t$.
\end{definition}

The supply side is the sum of total RES generation and the aggregate discharge of all storage players. The demand side is aggregate load, including all storage charging. Thus, both supply and demand are explicitly modeled to determine the market-clearing price. 

\begin{definition}\label{def_residual}
Let $\mathrm{RES}_t$ be renewable availability at hour $t$, and $(q^{\mathrm{ch},o}_t, q^{\mathrm{dis},o}_t)$ the aggregate rivals’ charge/discharge. The net supply offered to the day-ahead market is

$$q^{\mathrm{tot}}_t = \mathrm{RES}_t - q^{\mathrm{ch},p}_t + q^{\mathrm{dis},p}_t - q^{\mathrm{ch},o}_t + q^{\mathrm{dis},o}_t.$$

Given this net supply, let $V^{\max}_t$ denote the maximum cumulative demand volume at hour $t$ (the last step of the aggregate bid stack). The market \emph{cleared quantity} is
$q^{\mathrm{*}}_t = \min\{\,q^{\mathrm{tot}}_t, V^{\max}_t\}.$
Unmet demand (scarcity) and curtailed production (surplus) are
$u_t = \bigl(V^{\max}_t - q^{\mathrm{tot}}_t\bigr)^{+},\quad cur_t = \bigl(q^{\mathrm{tot}}_t - V^{\max}_t\bigr)^{+},$
so that
$q^{\mathrm{*}}_t = q^{\mathrm{tot}}_t - cur_t,\quad V^{\max}_t = q^{\mathrm{*}}_t + u_t,$ with $(x)^{+} = \max\{x,0\}$. In this framework, the state observed by player $p$, comprising other players' decisions and exogenous factors, is $\{q_{t}^{\mathrm{ch},o}, q_{t}^{\mathrm{dis},o}, \mathrm{RES}_t, \mathrm{vol}_t, \mathrm{pr}_t, e_{t-1}^{p} \}$.
\end{definition}

\begin{prop}\label{prop_profit}
Given \cref{ass_behavior} and \cref{ass_scope}, each storage player $p$'s profit is expressed linearly as price-action product: 
\begin{equation}\label{eq_prop}
\Pi_{t}^p=\sum_{i=1}^N \sum_{j=1}^D Q_i^p\, \mathrm{pr}_{tj} \left(w_{tij}^{\mathrm{dis},p} - w_{tij}^{\mathrm{ch},p} \right)
- OC_p \sum_{i=1}^N Q_i^p \left(z_{ti}^{\mathrm{dis},p} + z_{ti}^{\mathrm{ch},p} \right),
\end{equation}
where $w_{tij}^{\mathrm{dis},p}$ and $w_{tij}^{\mathrm{ch},p}$ are auxiliary binary variables defined as $z_{ti}^{\mathrm{dis},p}u_{tj}$ and $z_{ti}^{\mathrm{ch},p}u_{tj}$, respectively.
\end{prop}
\begin{prof}
The per-period profit of the storage player $p$ is the revenue from discharging minus the cost of charging, which can be expressed as 
\begin{equation}
\Pi^p_t = q^{\mathrm{dis},p}_t (\lambda_t - OC_p) -q^{\mathrm{ch},p}_t (\lambda_t + OC_p). 
\end{equation}
Taking the expression for charging and discharging from \cref{def_levels}, discretization of power levels and price from \cref{def_residual} and \cref{def_price}, respectively, profit of each storage player $p$ can be expressed as
\begin{equation}\label{OF}
\Pi^p_t = \sum_{i=1}^N Q_i^p\, z_{ti}^{\mathrm{dis},p} \left(\sum_{j=1}^D \mathrm{pr}_{tj} \, u_{tj} - OC_p \right) -\sum_{i=1}^N Q_i^p\, z_{ti}^{\mathrm{ch},p} \left(\sum_{j=1}^D \mathrm{pr}_{tj} \, u_{tj} + OC_p \right).
\end{equation}
Regrouping terms and applying charging and discharging decisions yield: 
\begin{subequations}\label{profit} 
\begin{align}
&\Pi^p_t = \sum_{i=1}^N \sum_{j=1}^D Q_i^p\, \mathrm{pr}_{tj} \, u_{tj} \left(z_{ti}^{\mathrm{dis},p} - z_{ti}^{\mathrm{ch},p} \right)
+ OC_p \sum_{i=1}^N Q_i^p \left(z_{ti}^{\mathrm{dis},p} + z_{ti}^{\mathrm{ch},p} \right) \\
&= \sum_{i=1}^N \sum_{j=1}^D Q_i^p\, \mathrm{pr}_{tj} \left(w_{tij}^{\mathrm{dis},p} - w_{tij}^{\mathrm{ch},p} \right)
- OC_p \sum_{i=1}^N Q_i^p \left(z_{ti}^{\mathrm{dis},p} + z_{ti}^{\mathrm{ch},p} \right)
\end{align}
\end{subequations}
The $w$ variables represent the product of a power level decision and a price-step selection. Although $w^{\mathrm{ch},p}_{tij}$ and $w^{\mathrm{dis},p}_{tij}$ are bilinear by definition, they can be linearized by imposing the following three additional constraints:
\begin{equation} \label{lin_w}
w_{tij}^{\mathrm{ch},p} \leq z_{ti}^{\mathrm{ch},p}, \qquad
w_{tij}^{\mathrm{ch},p} \leq u_{tj}, \qquad
w_{tij}^{\mathrm{ch},p} \ge z_{ti}^{\mathrm{ch},p} + u_{tj} - 1,
\end{equation}
and similarly for $w_{tij}^{\mathrm{dis},p}$. This ensures $w = 1$ if and only if both $z=1$ and $u=1$, and $w=0$ otherwise, exactly capturing the logical relationship $w = z \cdot u$. By including these linear constraints, the profit expression \eqref{profit} becomes fully linear and suitable for an MILP formulation. 
\end{prof}

The battery dynamics are modeled as follows. The energy level $e^p_t$ of player $p$’s battery evolves each period $t$ based on charging and discharging decisions:
\begin{equation}
e^p_1 = \alpha^{\mathrm{batt}} E_p^{\max} + \eta^p q_1^{\mathrm{ch}, p} - q_1^{\mathrm{dis},p} \label{eq_e1}
\end{equation}
\begin{equation}
e^p_t = e^p_{t-1} + \eta^p q_t^{\mathrm{ch}, p} - q_t^{\mathrm{dis},p} \quad \forall t, \label{eq_e_evol}
\end{equation}
where $\alpha^{\mathrm{batt}} E_p^{\max}$ is the initial energy (a fraction $\alpha^{\mathrm{batt}}\in[0,1]$ of capacity) and $\eta^p$ is the round-trip efficiency (assumed to affect charging only, without loss of generality). The state of charge is bounded by physical limits $0 \le e^p_t \le E_p^{\max}$. 

Finally, since the model optimizes operations over a single day, a constraint is introduced to ensure that the final energy level of the battery returns to (or remains close to) its initial value. This constraint mitigates boundary effects and prevents the optimizer from exhausting the battery purely to improve end-of-horizon profit. For a one-day benchmark, this is a standard way to represent periodic operation without imposing exact interday continuity. Without such a constraint, players would have an incentive to fully discharge their storage by the end of the time horizon to maximize short-term revenues. However, in a realistic setting with a daily (or other short-term) horizon, beginning the next period with an empty battery would likely be suboptimal. It is worth noting that such a constraint would not be appropriate for very short time horizons (e.g., a few hours), as there may be insufficient time to both charge and discharge the battery meaningfully \citep{nasrolahpour2016bidding,zhao2022strategic}.

Accordingly, rather than imposing exact equality, we require the final battery level to lie within a specified range around the initial level. This adjustment is necessary because charging and discharging power levels are discretized, whereas the battery energy level evolves continuously, especially under non-ideal efficiency conditions. Sensitivity to this terminal-state-of-charge assumption is reported in the Supplementary material (Table~S1 and Figure~S1).
\begin{equation}
\alpha^{\mathrm{batt}} E_p^{\max} (1 - \epsilon) \leq e_T^p \leq \alpha^{\mathrm{batt}} E_p^{\max} (1 + \epsilon) \label{eq_bound_eT}
\end{equation}
Expression (\ref{eq_ch_dis1}) ensures that at most one nonzero action is chosen per period (with zero indicating idle). This guarantees the player selects a single power level (and not a combination) in each hour:
\begin{equation}
\sum_{i=1}^N \left( z_{ti}^{\mathrm{ch},p} + z_{ti}^{\mathrm{dis},p} \right) \leq 1 \quad \forall t \label{eq_ch_dis1}
\end{equation}

\begin{definition}\label{def_price}
Let $u_{tj}\in\{0,1\}$ indicate the active demand step $j$ at hour $t$ with the discrete selection of the market-clearing price step $\sum_{j=1}^D u_{tj} = 1 \quad \forall t $. Then the market price is defined as
\begin{equation}\label{eq_price_discrete1}
\lambda_t=\sum_{j=1}^D \mathrm{pr}_{tj}\,u_{tj},
\end{equation}
where the interval consistency between $q^{\mathrm{tot}}_t$ and the active step is enforced either by the big-$M$ step-selection constraints detailed in the Supplementary material (Section~S9), or by the continuous block formulation \eqref{eq_alt_balance}–\eqref{eq_alt_b}.
\end{definition}

The market price $\lambda_t$ equals $\mathrm{pr}_{tj}$ exactly when the total dispatched energy $q_t^{\mathrm{tot}}$ lies within the cumulative-volume interval $[\mathrm{vol}_{t,j-1},\mathrm{vol}_{tj}]$ of the active step $j$. This interval-selection logic admits two equivalent linearizations. The first is a big-$M$ mixed-integer linear program (\textbf{Method 1}), which introduces the range indicators $u_{tj}^{+},u_{tj}^{-}$ and the product variables $w_{tij}^{\mathrm{ch},p},w_{tij}^{\mathrm{dis},p}$ to encode the active demand block. Because the big-$M$ linearization is a well-established technique for strategic market participation, its full derivation and the complete MILP program are reported in the Supplementary material (Section~S9).

It should be noted that while the big-$M$ formulation is commonly applied in Cournot games, big-$M$ constants can introduce numerical and computational challenges. If $M_t$ is chosen excessively large, the MILP relaxation becomes weak, potentially leading to slow convergence or even incorrect integer solutions due to tolerances. Thus, careful calibration of $M_t$ for each constraint is essential. For instance, $M_t$ can be set to the maximum feasible $q_t^{\mathrm{tot}}$ (e.g., total available supply plus demand at time $t$ when $u_{tj}=1$). However, tuning big-$M$ values proved to be challenging in models under many demand curve steps. To that end, we propose an alternative linearization of the bilinear products that eliminates big-$M$ constraints. The alternative method is not only equivalent to the MILP with a big-$M$ formulation, but it also eliminates the need to choose valid $M$ constants and typically yields better computational performance \citep{de2002price}.

The alternative method (\textbf{Method 2}) introduces continuous variables to directly model the portion of the demand curve served at the market-clearing price, thereby eliminating the need for $u^+_{tj}$, $u^-_{tj}$, and $w_{tij}^{p}$ variables and all big-$M$ constraints. We retain the same binary $u_{tj}$ from \eqref{eq_price_discrete1} to indicate the active price step, but introduce a continuous variable $b_{tj} \ge 0$ to represent the fraction of demand block $j$ that is served (filled) at time $t$. Hourly residual demand curves enable the precise formulation of the self-scheduling profit maximization problem that every price maker faces daily in a pool-based electricity market for energy.

\begin{prop}\label{prop_block}
Given the step-selection binary variable $u_{tj}$ and a continuous variable $0 \le b_{tj} \le q^{\max}_{tj} u_{tj}$, there exists a unique step $j^* \in D$ with $u_{tj^*}=1$ where the clearing quantity is 
\begin{equation}\label{crl_q}
q_t^{\mathrm{tot}}=\sum_{j\in D}\big(q^{\min}_{tj}u_{tj}+b_{tj}\big)
\end{equation}
where $q^{\min}_{tj}$ is the cumulative power volume (steps) from step 1 to $j-1$ for hour $t$.
\end{prop}
\begin{prof}
Let $\{\mathrm{vol}_{t,j}\}_{j=0}^{D}$ be the cumulative breakpoints with
$\mathrm{vol}_{t,0}=0$ and $\mathrm{vol}_{t,j}=\sum_{k=1}^{j} q^{\max}_{tk}$,
so that $q^{\min}_{tj}=\mathrm{vol}_{t,j-1}$ for each step $j\in D$.
By the feasibility of the continuous–block formulation, we have
$u_{tj}\in\{0,1\}$ with $\sum_{j\in D}u_{tj}=1$ and
$0\le b_{tj}\le q^{\max}_{tj}\,u_{tj}$.
Hence, there exists a unique index $j^\star\in D$ such that $u_{tj^\star}=1$
and $u_{tk}=0$ for all $k\neq j^\star$; the bound
$0\le b_{tk}\le q^{\max}_{tk}\,u_{tk}$ then implies $b_{tk}=0$ for $k\neq j^\star$.
Substituting this into \eqref{crl_q} gives
$$q_t^{\mathrm{tot}}
= \sum_{j\in D}\big(q^{\min}_{tj}u_{tj}+b_{tj}\big)
= q^{\min}_{t j^\star}+b_{t j^\star}.$$
Using $0\le b_{t j^\star}\le q^{\max}_{t j^\star}$ yields the bounds
$$q_t^{\mathrm{tot}}\in
\big[q^{\min}_{t j^\star},\,q^{\min}_{t j^\star}+q^{\max}_{t j^\star}\big]
=\big[\mathrm{vol}_{t,j^\star-1},\,\mathrm{vol}_{t,j^\star}\big].$$
Therefore, the clearing quantity lies \emph{on step $j^\star$ of the
price-quota curve}, at the point located a horizontal distance
$b_{t j^\star}$ to the right of the step’s left breakpoint $\mathrm{vol}_{t,j^\star-1}$.
No other step can contribute to $q_t^{\mathrm{tot}}$ because $u_{tk}=0$ implies
$b_{tk}=0$ for $k\neq j^\star$. Consequently, the clearing step is unique, and its
location on the curve is exactly identified by the interval
$[\mathrm{vol}_{t,j^\star-1},\,\mathrm{vol}_{t,j^\star}]$ and the offset $b_{t j^\star}$.
\end{prof}

The key constraints replacing the big-$M$ logic are:
\begin{subequations}\label{eq_alt}
\begin{align}
&q_t^{\mathrm{tot}} = \sum_{j=1}^D \Big( q_{tj}^{\min}u_{tj}+ b_{tj}\Big) \quad \forall t, \label{eq_alt_balance} \\
&0 \le b_{tj} \le q_{tj}^{\max} u_{tj}, \quad \forall t, j, \label{eq_alt_b}
\end{align}
\end{subequations}
where $q_{tj}^{\min}$ is the cumulative volume up to step $j-1$ (with $q_{t1}^{\min}=0$) and $q_{tj}^{\max}$ is the volume of demand in step $j$ (the increment $\mathrm{vol}_{tj} - \mathrm{vol}_{t,j-1}$). Constraint \eqref{eq_alt_balance} constructs the total dispatched quantity $q_t^{\mathrm{tot}}$ as the sum of: \textit{i)} all demand from higher-priced blocks fully served ($q_{tj}^{\min}$ for the active block $j$) and \textit{ii)} a partial quantity $b_{tj}$ from the active block $j$. Because by \eqref{eq_price_discrete1}, exactly one term in the sum \eqref{eq_alt_balance} will include a nonzero $q_{tj}^{\min}$, namely the active demand block. For that same block $j$, $b_{tj}$ can take a value up to the block’s size $q_{tj}^{\max}$, as enforced by \eqref{eq_alt_b}, to supply any remaining portion of that demand step. All other blocks $k \neq j$ have $u_{tk}=0$, forcing $b_{tk}=0$ in \eqref{eq_alt_b}. This construction guarantees that \textit{i)} if demand block $j$ is chosen ($u_{tj}=1$), then all higher price (lower volume) blocks $1,\dots,j-1$ are completely supplied (since $q_{tj}^{\min} = \mathrm{vol}_{t,j-1}$ is included in $q_t^{\mathrm{tot}}$), and block $j$ is supplied up to $b_{tj}$ (potentially partially); \textit{ii)} the total supply $q_t^{\mathrm{tot}}$ cannot exceed the maximum demand $\mathrm{vol}_{tD}$, because even if the last block $D$ is active ($u_{tD}=1$) we have $q_t^{\mathrm{tot}} = \mathrm{vol}_{t,D-1} + b_{tD} \le \mathrm{vol}_{t,D-1} + q_{tD}^{\max} = \mathrm{vol}_{tD}$. Thus, constraints \eqref{eq_alt_balance}–\eqref{eq_alt_b} embed the piecewise-linear demand curve into the model.

Under this alternative formulation, the objective function remains total-profit maximization, with only the revenue term represented through $\mathrm{pr}_{tj}(b_{tj}+u_{tj}q_{tj}^{\min}) $. In practice, revenue in each period can be expressed using the market-clearing price and the cleared quantity. However, since all players receive the uniform market price $\lambda_t$ for the energy they sell, player $p$’s revenue can be written as $\lambda_t q_t^{\mathrm{dis},p}$ (and its charging cost as $\lambda_t q_t^{\mathrm{ch},p}$), consistent with the profit definition given earlier (\cref{prop_profit}). All the operational constraints for battery dynamics and power limits remain unchanged in this formulation. In other words, the only differences lie in the market-clearing constraints \eqref{eq_alt_balance}–\eqref{eq_alt_b}, which replace the big-$M$ step-selection constraints (Supplementary material, Section~S9) and eliminate the need for the $w$ variables in the profit calculation. We therefore solve the following alternative MILP for each player $p$:
\begin{subequations}\label{alter}
\begin{align}
&\max_{\chi^p_{\mathrm{alt}}} \sum_{t}\left[\sum_{j} \mathrm{pr}_{tj}(b_{tj}+u_{tj}q_{tj}^{\min}) 
- OC_p \sum_{i=1}^N Q_i^p \left(z_{ti}^{\mathrm{dis},p} + z_{ti}^{\mathrm{ch},p} \right) \right]\qquad\\
&\text{s.t.} \nonumber\\
&q_t^{\mathrm{tot}} = \sum_{j=1}^D \Big( q_{tj}^{\min}u_{tj}+ b_{tj}\Big) \quad \forall t, \\
&0 \le b_{tj} \le q_{tj}^{\max} u_{tj}, \quad \forall t, j,\\
& 0 \leq e_t^p \leq E_p^{\max} \quad \forall t,p \\
& \alpha^{\mathrm{batt}} E_p^{\max} (1 - \epsilon) \leq e_T^p \leq \alpha^{\mathrm{batt}} E_p^{\max} (1 + \epsilon) \\
& e_1^p = \alpha^{\mathrm{batt}} E_p^{\max} + \eta^p q_1^{\mathrm{ch},p} - q_1^{\mathrm{dis},p} \\
& e_t^p = e_{t-1}^p + \eta^p q_t^{\mathrm{ch},p} - q_t^{\mathrm{dis},p} \quad \forall t \\
& \sum_{i=1}^N \left( z_{ti}^{\mathrm{ch},p} + z_{ti}^{\mathrm{dis},p} \right) \leq 1 \quad \forall t \\
& \sum_{j=1}^D u_{tj} = 1 \quad \forall t,
\end{align}
\end{subequations}
where $\chi^p_{\mathrm{alt}} = \{z^{\mathrm{ch},p}, z^{\mathrm{dis},p}, e^p, u, b\}$ is the set of decision variables. By \cref{prop_block}, Method 2 is mathematically equivalent to the big-$M$ model in the sense that it yields the same outcomes for each player and thus the same Nash equilibria when players interact. However, it offers several advantages. \textit{i)} It uses fewer binary variables: the $2D$ auxiliary binaries $u^+_{tj},u^-_{tj}$ are no longer needed, nor are the $2\times N\times D$ product binaries $w^{\mathrm{ch},p}_{tij},\,w^{\mathrm{dis},p}_{tij}$ (i.e., two actions $\times$ $N$ levels $\times$ $D$ price blocks per $(t,p)$). This reduction can significantly improve computational tractability, especially as the number of price steps $D$ grows. \textit{ii)} The formulation avoids introducing very large coefficients into the constraint matrix, thereby improving the linear programming relaxation. Our subsequent simulations and analyses employ Method 2, because the big-$M$ approach requires case-by-case fine-tuning.

\subsubsection{Solution approach}
The optimization problem \eqref{alter} (equivalently, the big-$M$ program in the Supplementary material) can be solved independently for each player, where each player makes assumptions about the behavior of the others. The search for a Nash equilibrium is performed iteratively. Initially, the game is run with all players assuming that the others adopt a zero-action strategy (i.e., no charging or discharging over the entire time horizon), which provides a simple and feasible starting point for the best-response updates. The output of each run consists of the player's charging and discharging decisions at every time step. The collective state of the system is defined by the set of all individual player strategies. In the second iteration, each player updates its strategy based on the strategies obtained from the first round for the other players. At each iteration, the new system state is compared with that of the previous one. This process continues until the strategy profile of a round stabilizes. In implementation, we monitor both the maximum change in players' strategies and the maximum relative change in players' profits across successive passes. Convergence is declared when both quantities fall below preset tolerances. This approach is similar to the diagonalization algorithm (Gauss-Seidel-type method) \citep{hu2007using} for computing Nash equilibria in electricity markets. Furthermore, since power outputs are modeled using discrete variables, the number of feasible strategies for each player is finite, which makes cycling easy to detect and provides a practical basis for the stopping rule.

\begin{algorithm}[H]
	\small
\caption{Iterative search for Nash equilibrium}
\label{alg_nash_equilibrium}
\begin{algorithmic}[1]
\Procedure{NashEquilibrium}{$q^{\mathrm{ch, o}}_{\mathrm{ini}}, q^{\mathrm{dis,o}}_{\mathrm{ini}}, \mathrm{state}_{\mathrm{ini}}$}
\State Initialize $\texttt{state\_history, state} \gets \{\}, \{\}$
\For{$p \in P$}
\State $\mathrm{state}[p] \gets \Call{ProfitMaximisation}{q^{\mathrm{ch,o}}_{\mathrm{ini}}, q^{\mathrm{dis,o}}_{\mathrm{ini}}, \mathrm{state}_{\mathrm{ini}}}$ solving (\ref{alter})
\EndFor
\State $\texttt{state\_sys} \gets [\mathrm{state}[p]\quad \forall p \in P]$
\State $k \gets 1$
\While{$\texttt{state\_sys} \notin \texttt{state\_history}$}
\State $\texttt{state\_history}[k] \gets \texttt{state\_sys}$
\For{$\forall p \in P$}
\State $q^{\mathrm{ch,o}} \gets \sum_{i \neq p} \mathrm{state}[o][1]$
\State $q^{\mathrm{dis,o}} \gets \sum_{i \neq p} \mathrm{state}[o][2]$
\State $\mathrm{state}[p] \gets \Call{ProfitMaximisation}{\mathrm{state}, q^{\mathrm{ch,o}}, q^{\mathrm{dis,o}}}$ 
\EndFor
\State $\texttt{state\_sys} \gets [\mathrm{state}[p]\quad \forall p \in P]$
\State $k \gets k + 1$
\EndWhile
\State \textbf{return} $\texttt{state\_sys}$ 
\EndProcedure
\end{algorithmic}
\end{algorithm}

Algorithm \ref{alg_nash_equilibrium} outlines the iterative search for a Nash equilibrium, progressing through a series of strategy updates until the system reaches convergence. Step $1$ initializes the system by assuming all players follow a zero-action strategy, i.e., no charging or discharging over the entire time horizon. Step $2$ solves the individual profit-maximization problem for each player $p$, given the assumed actions of the other players. This yields each player’s initial strategy (their charging and discharging schedule), and the overall system state is stored as a collection of these strategies. Step $3$ begins the iterative update loop, storing the current system state in a history log. In Step $4$, each player, in turn, updates its strategy assuming the most recent strategies of all other players remain fixed. The player then solves its optimization problem and adopts the resulting best-response strategy. In Step $5$, the algorithm checks whether the newly computed system state matches any previously recorded state. If a match is found, convergence is reached, and the process terminates. If not, Step $6$ logs the new state, and the algorithm returns to Step $4$ for another iteration. Finally, Step $7$ returns the converged strategy profile, which constitutes a Nash equilibrium, as no player can unilaterally improve their payoff by deviating.

We assessed the robustness of the diagonalization by initializing from multiple random strategy profiles. In the reported experiments, the best-response routine uses tolerances of $\epsilon_q=10^{-5}$ MW for the maximum strategy change and $\epsilon_\pi=10^{-6}$ for the maximum relative profit change, together with a requirement that both conditions hold for two consecutive passes and a maximum of 50 iterations. All scenarios converged within 21 iterations. To investigate the existence of multiple equilibria, we employ a multi-start approach, running the solver from various initial points to discard low-quality optima and isolate the best solution \cite{abate2025dynamic}. Applying this method to the two-player case reveals significant sensitivity to initial conditions. Notably, the equilibrium structure differs by season: winter converges on a dominant, high-welfare equilibrium, whereas summer admits several equilibria with comparable welfare levels. For three or more operators the best-response iteration converges rapidly and monotonically without cycling. Detailed multi-start and convergence evidence is reported in the Supplementary material (Section S7 and Figures S6–S9).

For the big-$M$ formulation, $M_t$ values are chosen as tight, time-varying upper bounds derived from the maximum feasible cleared volume at hour $t$, rather than through ad hoc calibration. The model is solved with Gurobi 12.0 on an Intel Core i5-135U (1.60 GHz) using Python 3.12.7, and a relative optimality tolerance of $10^{-6}$. Method 2 is used for the main analysis because it was more numerically stable and computationally efficient on the full scenario grid.

\subsection{System level problem formulation} \label{model_sw}

Player bidding strategies influence the market-clearing outcome and the amount of demand served. In an oligopolistic setting, such strategic behavior often leads to inefficiencies and potential welfare losses. To quantify these losses, we consider a centralized benchmark scenario in which a system operator optimally coordinates the operation of all players, aiming to maximize overall social welfare. The main assumptions of the benchmark are: \textit{i}) all storage players are centrally coordinated by the system operator to achieve the socially optimal outcome, and \textit{ii}) the system operator has complete information on players' technical parameters, the market demand curve, and renewable energy production. Consumer surplus represents the difference between consumers' willingness to pay and the market-clearing price. In this setting, it corresponds to the area under the demand curve and above the market-clearing price.

To model this, let $y_{tj}$ be a binary variable that takes the value $1$ if and only if the bid price $\mathrm{pr}_{tj}$ exceeds the market-clearing price. This condition is equivalent to the associated cumulative volume bid $\mathrm{vol}_{tj}$ being less than the market-clearing volume.
\begin{align}
&\sum_{k=1}^D u_{tk} = 1, \quad u_{tk}\in\{0,1\}, \quad
u_{tk}=1 \Longleftrightarrow\; \mathrm{pr}_{tk}\ge\lambda_t,\\[6pt]
&y_{tj} = \sum_{k=j+1}^D u_{tk},\quad y_{tj}\in\{0,1\}.
\end{align}
The corresponding optimization problem is then defined as follows, with the key distinction from the strategic Method-2 program \eqref{alter} being the objective function:
\begin{subequations}
\begin{align}
& \max_{\boldsymbol{\chi^{\mathrm{SO}}}} \quad \sum_{t=1}^{T} (\mathrm{CS}_t + \mathrm{PS}_t)
\label{eq_objective_sw}\\
& \text{s.t.}\nonumber\\
& \text{\eqref{eq_e1}--\eqref{eq_ch_dis1}, \eqref{eq_price_discrete1}, \eqref{eq_alt_balance}--\eqref{eq_alt_b}},
\end{align}
\end{subequations}
where $\chi^{\mathrm{SO}} = \{z^{\mathrm{ch}}, z^{\mathrm{dis}}, e, u, b, y\}$ is the set of decision variables for the social planner's problem, including all control and binary variables required to determine the welfare-maximizing allocation. The objective is to maximize total social welfare, defined as the sum of consumer surplus ($\mathrm{CS}$) and producer surplus ($\mathrm{PS}$). Producer surplus, which includes the combined profits of all storage players and the surplus from renewable energy sources, is given by:
\begin{equation}
\mathrm{PS}_t = \sum_{p=1}^P \Pi^p_t + \Pi^{\mathrm{RES}}_t
\end{equation}
where the first term $\Pi^p_t$ is the period profit of player $p$, and the second term is the renewable energy producer surplus, explicitly calculated as the revenue from selling the renewable output at the market price ($\Pi^{\mathrm{RES}}_t = \lambda_t \mathrm{RES}_t$). Thus, $\mathrm{PS}$ captures the surplus of both storage players and renewable generators.

On the other hand, consumer surplus ($\mathrm{CS}$) at hour $t$ can be computed as:
\begin{equation}
\mathrm{CS}_t = \sum_{k=1}^{D} \big(\mathrm{pr}_{tk}-\lambda_t\big)\Delta \mathrm{vol}_{tk}\ y_{tk},
\qquad \Delta \mathrm{vol}_{tk}:=\mathrm{vol}_{tk}-\mathrm{vol}_{t,k-1}, \quad \mathrm{vol}_{t0}=0.
\end{equation}
Equivalently:
\begin{equation}
\mathrm{CS}_t =\ \sum_{k=1}^{D-1} \big(\mathrm{pr}_{tk}-\mathrm{pr}_{t,k+1}\big) \mathrm{vol}_{tk} y_{tk}.
\end{equation}
The indicator $y_{tk}=\sum_{m=k+1}^{D} u_{tm} = 1$ if and only if the bid price $\mathrm{pr}_{tk}$ strictly exceeds the market-clearing price $\lambda_t$; hence, only demand blocks strictly above $\lambda_t$ contribute to $\mathrm{CS}_t$. The marginal block has $\mathrm{pr}_{tj^*}=\lambda_t$ and contributes zero, and blocks below the clearing price do not contribute.

\section{Case study and simulation results}\label{design}

This section presents a detailed case study based on Denmark's projected 2030 renewable electricity system and evaluates the model's results under various scenarios. We first describe the data and scaling methodology for the 2030 scenario, including the renewable generation profiles and aggregate demand curves for representative winter and summer days. We then outline the simulation framework, followed by an analysis of market outcomes across different levels of storage competition and capacity scaling. Results for the winter and summer representative days are presented in parallel to analyze seasonal differences in market dynamics.

\subsection{Data scaling and simulation setup}\label{data}
We use 2024 hourly time-series data: (\textit{i}) capacity-factor profiles from Energi Data Service \citep{energidataservice2025} and (\textit{ii}) aggregated day-ahead bid stacks from Nord Pool (DK1) \citep{nordpool2025}. These data initialize the renewable and residual-demand blocks through a representative-day procedure and underpin the price-quantity steps used in market clearing. Because the model is designed as a stylized short-run benchmark, representative days are used to capture typical seasonal operating conditions while keeping the scenario design transparent and computationally tractable. Targeted robustness checks on representative-day choice and block granularity are summarized below and documented in detail in the Supplementary material Section~S5.

\subsubsection{Denmark 2030 renewable energy}
The Danish government has committed to phasing out fossil fuels entirely, aiming for a 100\% renewable electricity generation by 2030. According to projections from the Danish Energy Agency (DEA) \citep{DEA}, the generation mix will be dominated by wind power and solar photovoltaics (PV). This mix is expected to meet the projected annual electricity demand of 38 TWh in 2030, with approximately 75\% coming from wind, 12\% from solar PV, and 13\% from bioenergy. Because bioenergy is dispatchable and can act as a flexible baseload resource, it is excluded from our modeling to focus exclusively on the integration of variable renewable energy sources.

Hourly capacity factor (CF) time series for offshore wind, onshore wind, and solar PV in the DK1 bidding zone are obtained from the Danish Energy Data Service \citep{energidataservice2025}. To capture seasonal differences, we apply $k$-medoids clustering separately to winter and summer CF profiles and retain the medoid of the largest seasonal cluster as the model's \textit{typical day}. This is intended to represent a central seasonal operating pattern rather than an extreme realization. To assess the sensitivity of the results to this design choice, alternative representative days are analyzed in the Supplementary material (Section~S5, Table~S4 and Figures~S3--S4). Each representative day $\mathcal{D}_d$ is defined as

$$\text{Day}_d = \left\{ \left(cf_{\mathrm{solar},t}, cf_{\mathrm{offshore},t}, cf_{\mathrm{onshore},t} \right) \right\}_{t=0}^{23},$$

where $cf_{i,t}$ is the fraction of installed capacity for technology $i$ at hour $t$.

The representative-day CFs are then multiplied by the DEA’s 2030 installed capacity projections, yielding hourly MW profiles for winter and summer. This scales hourly RES generation to the projected 2030 system while preserving observed intraday renewable patterns. The resulting profiles provide stylized but data-grounded seasonal scenarios for analyzing storage competition under high renewable penetration.

\subsubsection{Demand curve construction}
To represent the demand side, we construct hourly demand curves from Nord Pool day-ahead data for DK1. Daily bid stacks are clustered into \emph{Winter} and \emph{Summer} groups via $k$-medoids, and the medoid of the largest cluster is used as the representative day for each season. Each day $d$ consists of 24 hourly price--quantity blocks,

$$\text{Day}_d = \left[\left( \text{Aggregated volume bids at time } t, \text{Aggregated price bids at time } t \right) \right]_{t = 0}^{23}.$$

Because the number of submitted bids differs across days and hours, we standardize the curves by approximating each hourly step function with a common number of blocks. Operationally, bids are ordered by price, cumulative volume is computed, and the original curve is compressed into $D$ representative blocks. This preserves the stepwise structure of the bid stack while ensuring a common dimensionality across hours and scenarios. The baseline uses $D=20$, which provides a practical compromise between fidelity to the original step curve and numerical tractability. Sensitivity to block granularity is reported in the Supplementary material (Table~S4 and Figure~S3).

In the aggregate bid curves, some hours contain negative-price blocks. In present-day electricity markets, negative offers arise primarily from four institutional drivers: thermal must-run conditions (start-up and shut-down costs that make short-term losses cheaper than cycling), production-linked support schemes (feed-in premia, production tax credits, and green-certificate revenues that let generators remain profitable even when the wholesale price is negative), CHP heat-demand couplings, and must-take contractual arrangements such as priority-dispatch rules or legacy power-purchase agreements. None of these drivers is present in the stylized 100\% RES, subsidy-free, energy-only market modeled in this paper. RES generators in the model face zero marginal fuel cost, receive no per-MWh subsidy, have no heat-side obligation, and can curtail instantaneously and costlessly. For such a generator, submitting a negative offer is strictly dominated by curtailment: curtailment yields zero revenue at zero cost, whereas a negative offer would yield negative revenue at zero cost. The profit-maximizing supply offer is therefore bounded below at zero, and negative offers are accordingly absent from the market setting analyzed in this paper.

\subsubsection{Storage sizing and other parameters}

We construct the residual demand and storage parameters directly from the scaled representative-day profiles. Below is the procedure for calculating the residual demand curves. Let $\mathrm{RES}_t$ denote renewable generation and $\mathrm{D}_t$ the total electricity demand at hour $t\in\{1,\dots,24\}$. The \emph{residual demand} is the net shortfall relative to renewables: $R_t = \mathrm{D}_t - \mathrm{RES}_t,$
where $R_t>0$ indicates a \emph{deficit} (renewables are insufficient) and $R_t<0$ a \emph{surplus}.

To translate hourly imbalances into a storage requirement, we correct $R_t$ for round-trip efficiency $\eta\in(0,1]$. To size storage compactly, we adjust deficits for a round-trip efficiency $(\eta)$ by setting $(R'_t=R_t\eta)$ if $(R_t>0)$ and $(R'_t=R_t)$ otherwise. Given the cumulative residual demand $(\mathrm{CR}_t=\sum_{l=1}^{t} R'_l)$, the local storage level is $(X_t=\mathrm{CR}_t-\min_{1\le i\le t}\mathrm{CR}_i)$. The required energy capacity is the maximum excursion $(E^{\max}=\max_t X_t)$. 
We partition total energy and power using fixed shares $(w_1,\dots,w_n)$ such that $(\sum_p w_p=1)$ (e.g., for $(n=2): ({1/3,2/3})$; for $(n=4): ({0.1,0.2,0.3,0.4})$, and so on). For player $(p)$, we set $(E^{\max}_p = w_p E^{\max})$ and $(Q^{\max}_p=\max\{1,\lfloor w_p\,Q^{\max}\rfloor\})$ (enforcing at least (1) MW if $(w_p>0))$. The discharge offer set is a strictly positive $(N)$-point grid, $(\mathcal{Q}_p={Q^{\max}_p/N,2Q^{\max}_p/N,\dots,Q^{\max}_p})$.

\subsection{Simulation design and result presentation}
We define the number of active storage operators $N$ and specify the corresponding market structure.  The analyzed configurations include \textit{Monopoly} (single operator), \textit{Duopoly} (two operators), and \textit{Oligopoly} (eight operators). Sensitivity analyses focus on levels of competition, storage efficiency, operational costs, and other key parameters. The problem is also solved from the social planner's perspective to establish an efficiency benchmark devoid of strategic withholding. Additional cases examine the impact of varying $N$ on overall market welfare. 
The \emph{seasonal simulations} are conducted for representative \textit{winter} and \textit{summer} days, enabling the analysis to capture seasonal variations in RES availability, demand patterns, and storage behavior. Simulation results are examined based on market-clearing prices (MCP), storage charging/discharging schedules, state-of-charge (SoC) trajectories, residual demand after RES and storage dispatch, unmet demand, and production curtailment.

\subsubsection{Monopoly (one storage operator)}

\begin{figure}[htbp]
\centering
\includegraphics[width=0.80\textwidth]{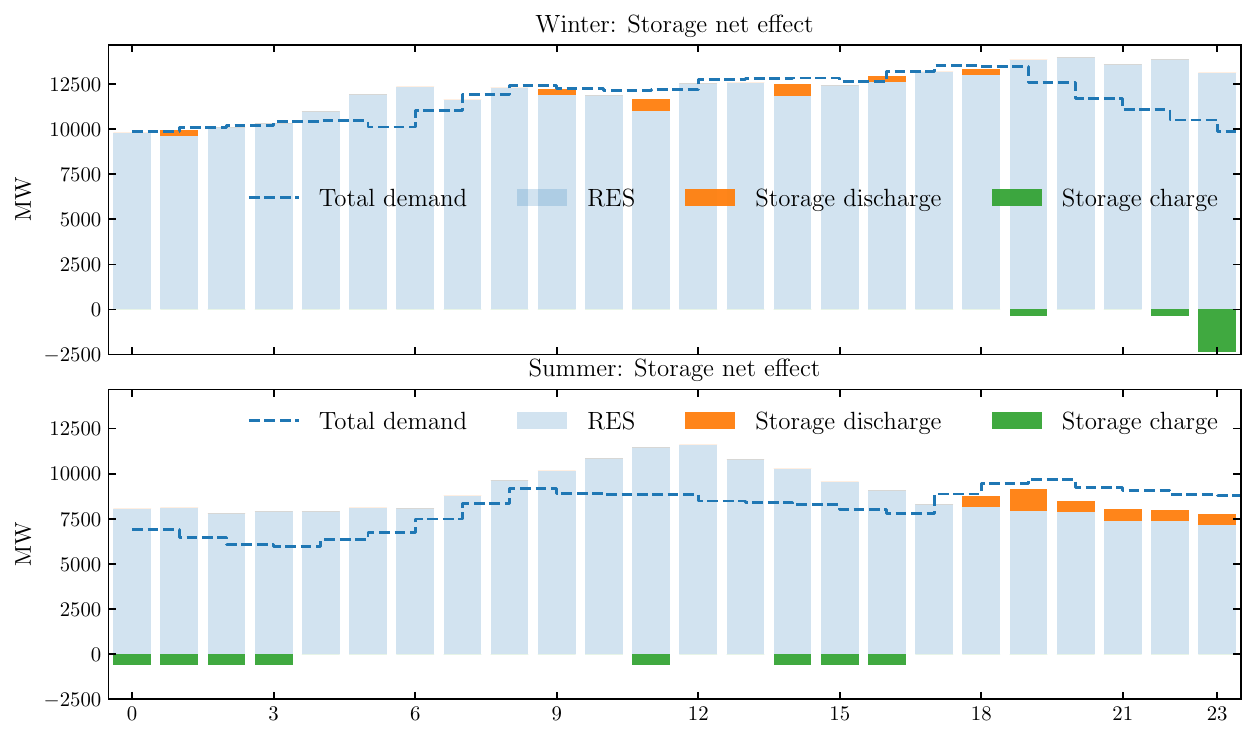}
\caption{Net storage effect for a representative winter and summer day with one storage operator (baseline capacity $\theta=1.0$, $D=20$ demand blocks).}
\label{net_1_storage}
\end{figure}

\begin{figure}[H]
\centering
\includegraphics[width=0.8\textwidth]{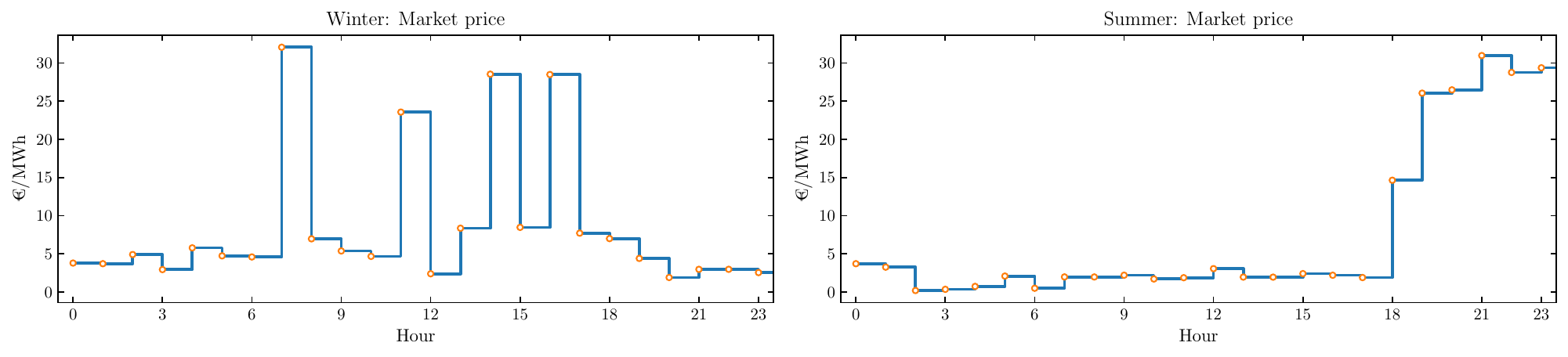}\\
\includegraphics[width=0.8\textwidth]{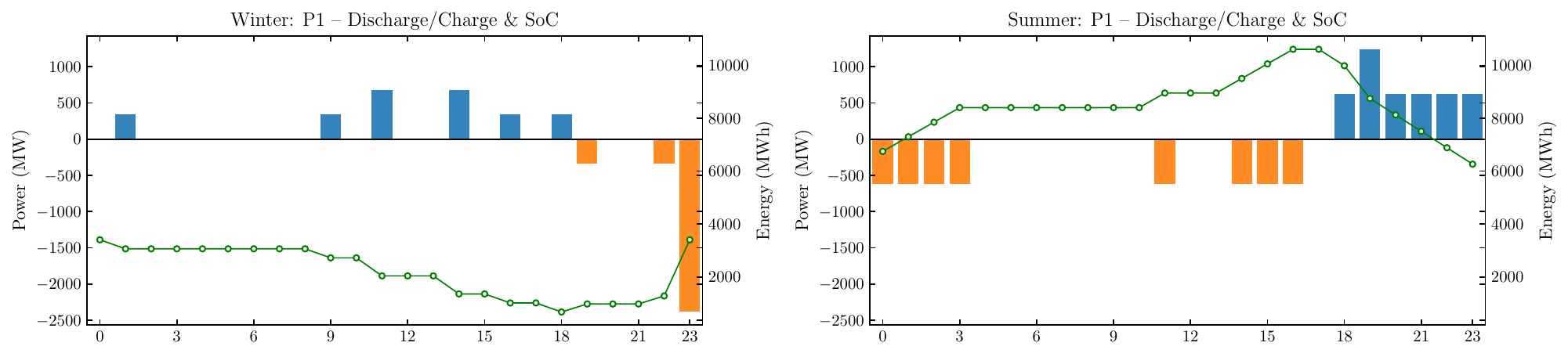}
\caption{Market-clearing prices, discharging (blue), charging (orange), and SoC (green) for a representative winter and summer day with one storage operator.}
\label{MCP_SoC}
\end{figure}
Figure \ref{net_1_storage} illustrates the role of a single storage operator in balancing supply and demand in the electricity market. The operator discharges during hours of RES shortage and charges during periods of surplus. Due to the specific demand and RES profiles of the winter day, the charging and discharging pattern appears less consistent. In contrast, during the summer day, discharging occurs exclusively during nighttime hours when shortages arise from the lack of solar PV generation, while charging takes place for seven hours between midnight and 19:00 when excess power is abundant. However, there are many hours during which no charging/discharging activity occurs, even though power surpluses or shortages exist. This behavior stems from the combination of the storage player's strategic profit-maximizing decisions and physical battery constraints.

Figure \ref{MCP_SoC} presents the maximum MCP and the storage operator's charging and discharging schedule. The MCP is about \EUR{}30/MWh with just one operator.\footnote{Prices around \EUR{}30/MWh arise from the empirical stepwise inverse-demand. A tiny \textit{necessity} block priced up to the cap is fully covered by exogenous RES, so the marginal unit clears on low-willingness-to-pay (WTP) steps (\(\approx\)\EUR{}30–80/MWh). Any remaining volume lies to the right of that step and is recorded as unserved, not priced at the cap. For a detailed analysis of actual scarcity pricing, see Section \ref{sec_scarcity-real}.} In summer, despite sustained surplus conditions throughout most hours of the day, the MCP rises slightly in some periods due to the operator's charging activity. The right axis of Figure \ref{MCP_SoC} illustrates the operator's charging decisions (MWh), which are expectedly higher on summer days, reflecting the greater availability of surplus generation.

\subsubsection{Duopoly (two storage operators)}
\begin{figure}[htbp]
\centering
\includegraphics[width=0.9\textwidth]{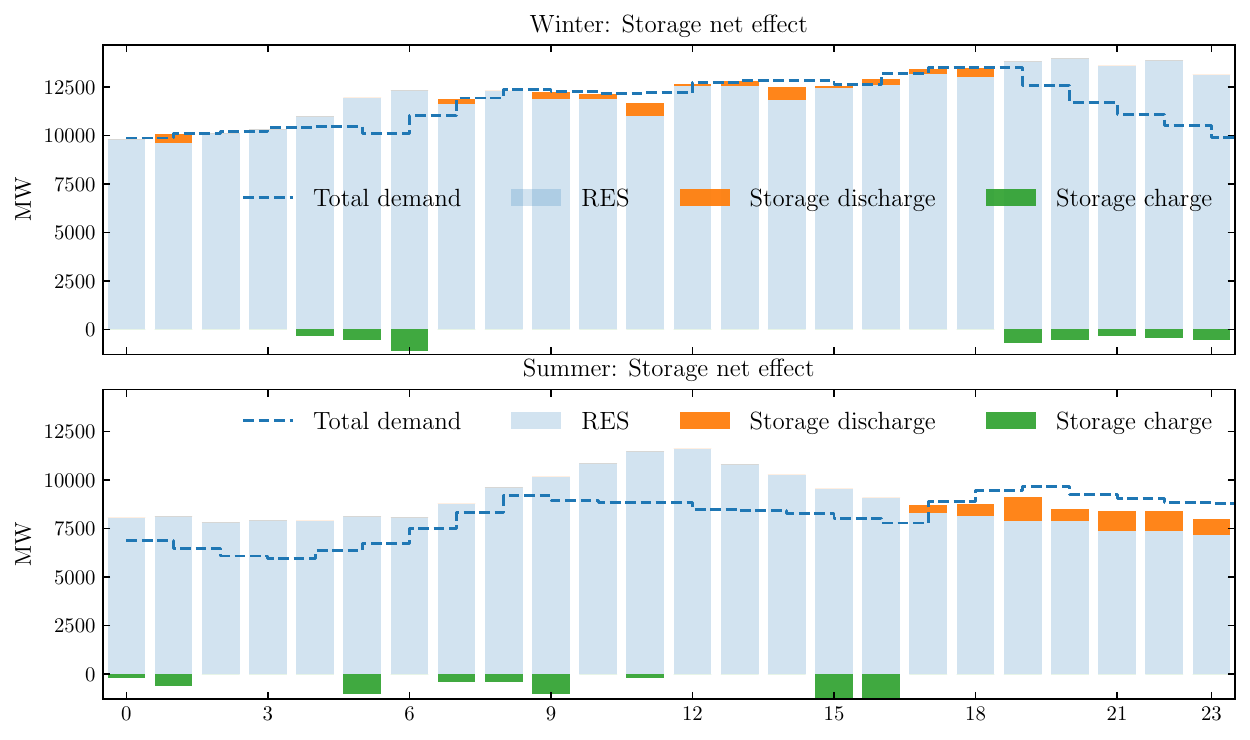}
\caption{Net storage effect for a representative winter and summer day under Duopoly.}
\label{net_2_storage}
\end{figure}

Figure \ref{net_2_storage} illustrates the net effect of storage with two players. In this scenario, both winter and summer days exhibit an increased number of charging and discharging hours. For instance, compared to the monopoly case on the winter day, where there are only three hours of charging, the Duopoly case shows eight hours of charging (in the early morning and late evening). Consequently, the duration of unmet demand during the winter day is noticeably reduced. This demonstrates that competition can effectively reduce electricity prices and unmet demand across both winter and summer days.

\begin{figure}[htbp]
\centering
\includegraphics[width=.9\textwidth]{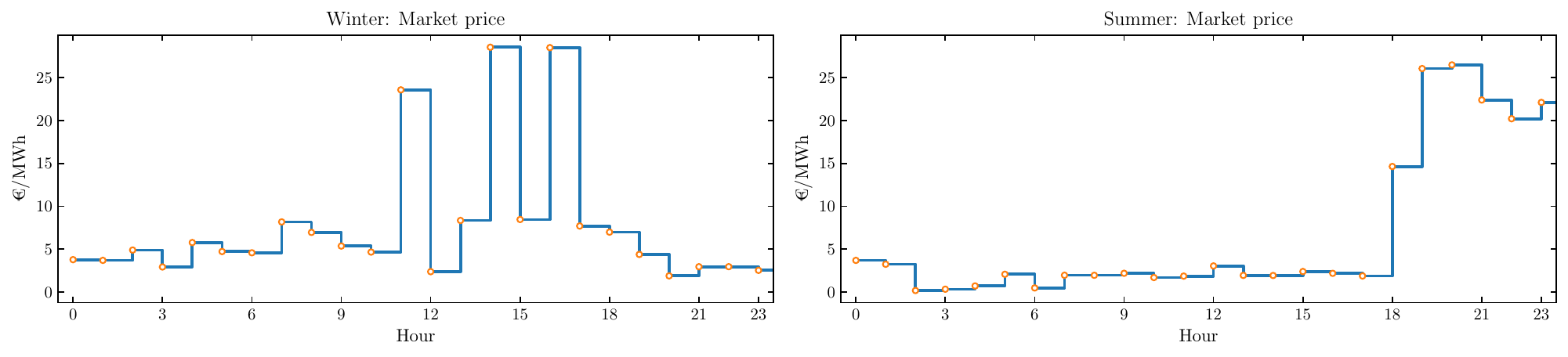}\\
\includegraphics[width=.9\textwidth]{compare_player_1_1p.pdf}\\
\includegraphics[width=.9\textwidth]{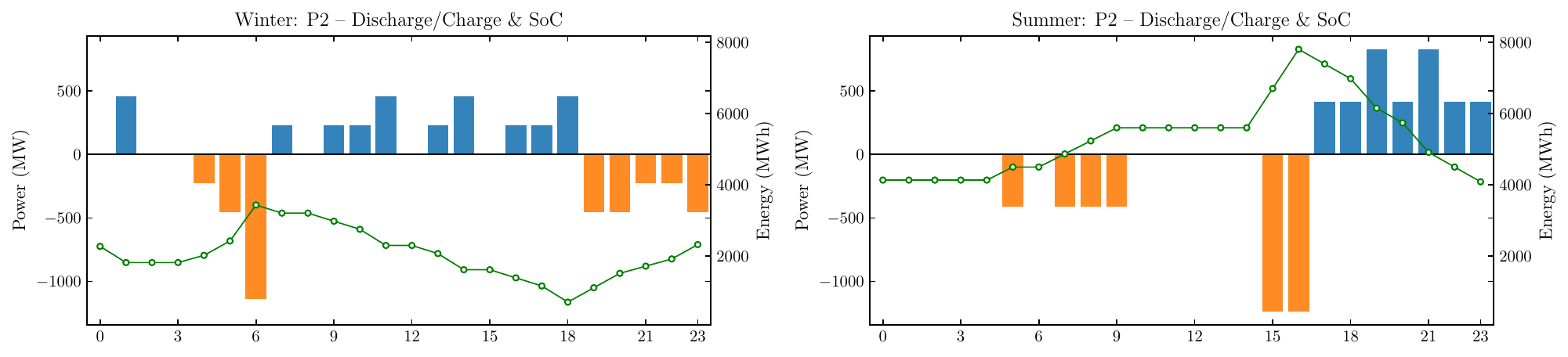}\\
\caption{Market-clearing prices, and charging (blue), discharging (orange), and SoC (green) for representative winter and summer days under Duopoly.}
\label{MCP_SoC_2}
\end{figure}
Figure \ref{MCP_SoC_2} presents electricity prices alongside each player’s charging/discharging profiles and SoC. The peak price decreases to \EUR{}25/MWh on both days, whereas the lowest price during RES-surplus hours increases slightly compared to the Monopoly case. 

Detailed per-player storage metrics for the Duopoly case (maximum charge and discharge, SoC range, throughput, profit, and cycling hours) are reported in Section~S10 of the Supplementary material. Player~2, which holds two-thirds of the allocated capacity, dominates across all metrics, earning a total profit of \EUR{}$110,741.51$ against \EUR{}$59,326.93$ for Player~1.

\subsubsection{Oligopolistic competition}
This section examines the impact of oligopolistic competition among storage players on market outcomes. In this configuration, we solve the model with eight storage players, distributing the fixed total storage capacity among them using different probability weights.

\begin{figure}[H]
\centering
\includegraphics[width=0.8\textwidth]{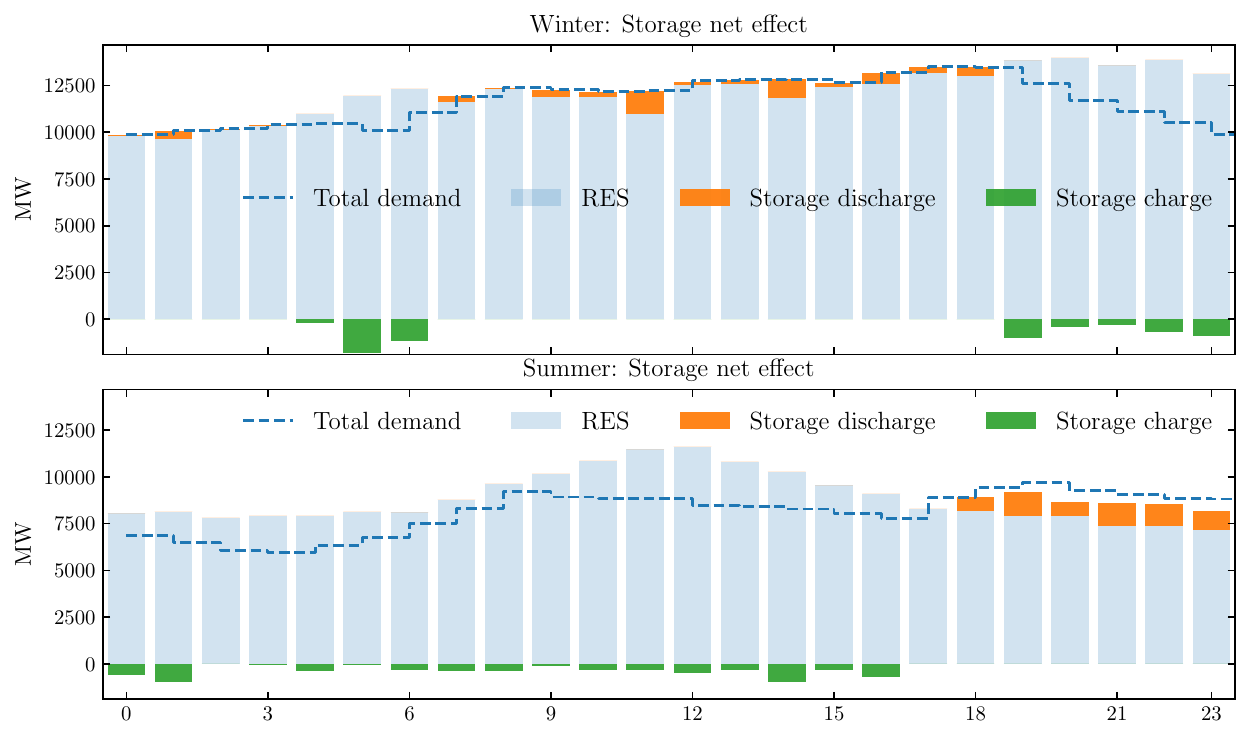}
\caption{Net storage effect for a representative winter and summer day under Oligopoly (eight storage operators).}
\label{net_8_storage}
\end{figure}

\begin{figure}[htbp]
\centering
\includegraphics[width=0.85\textwidth]{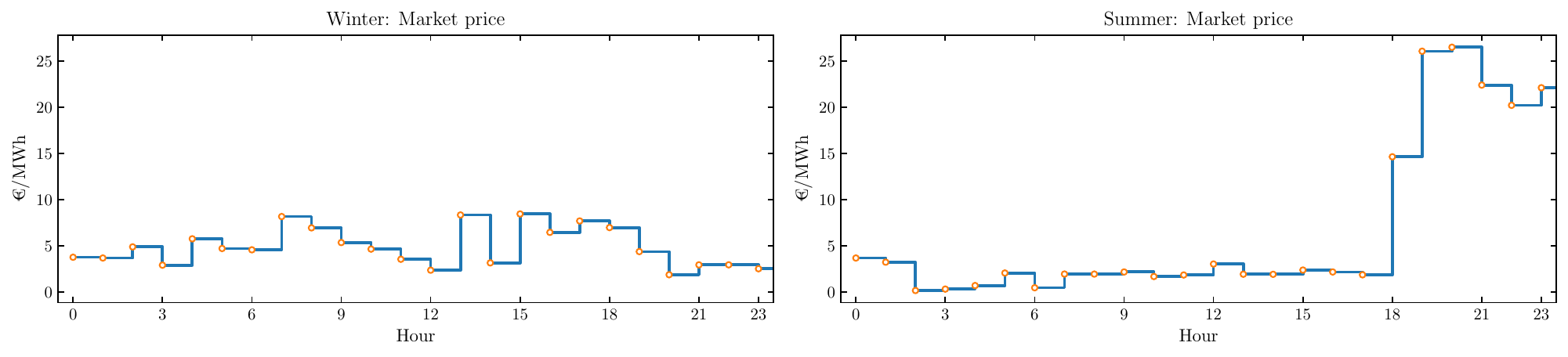}\\
\includegraphics[width=0.85\textwidth]{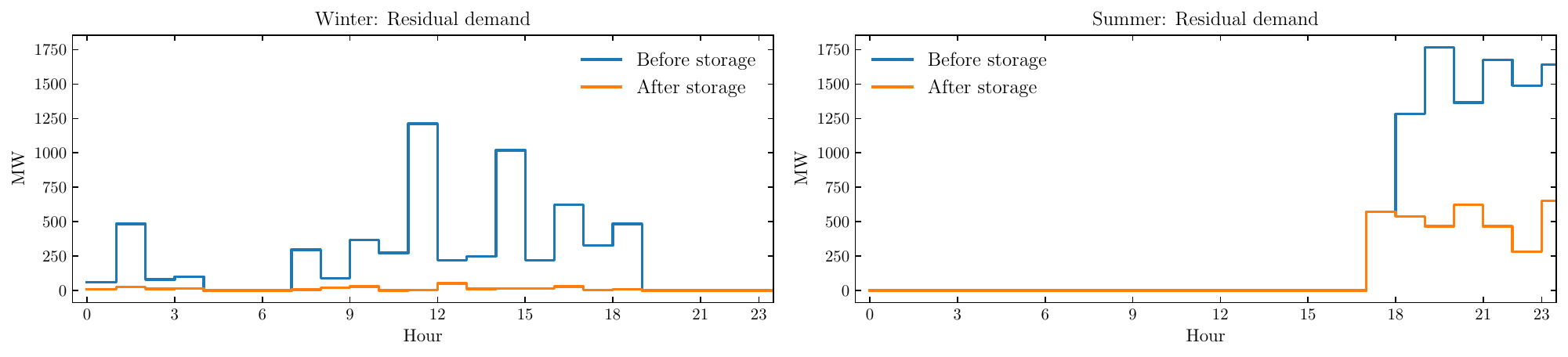}
\caption{Market-clearing prices and residual demand before and after the game for a representative winter and summer day under Oligopoly.}
\label{MCP_SoC_8}
\end{figure}

Per-player operational results under the oligopoly are compared in Section~S10 of the Supplementary material. Due to the availability of surplus energy, the winter day experiences a lower average MCP and, consequently, lower profits for all players compared to the summer day. In winter, both total profit and per-player profit decline as competition intensifies: without extended periods of surplus or deep unmet demand, firms merely split a limited arbitrage opportunity. In summer, by contrast, competition raises total producer surplus by driving down unmet demand and curtailment, even though each firm’s profit remains well below the monopoly level. In both seasons, the volume (utilization) of charging and discharging tends to increase with the number of competitors.

Figure \ref{net_8_storage} further shows that under an 8-player oligopoly, unmet demand is nearly eliminated, with the net storage effect approaching socially optimal levels. This is corroborated by Figure \ref{MCP_SoC_8}, which illustrates the residual demand patterns before and after storage dispatch. These seasonal differences in storage profitability, combined with variations in charging and discharging behavior, directly influence market-clearing prices, residual demand patterns, and ultimately overall welfare. Figure \ref{MCP_SoC_8} also demonstrates that in winter, electricity prices remain below \EUR{}5/MWh for most hours, whereas in summer they can reach as high as \EUR{}25/MWh, particularly during the evening peaks from 18:00 to 24:00.\footnote{Even under scarcity (i.e., when some demand is unmet), the MCP does not necessarily reach the maximum administrative price cap (\EUR{}4,000/MWh). Only the portion of demand willing to pay at or below the cap (e.g., 3,000 MW) clears, while demand above that level is curtailed or shiftable and therefore does not bid at higher prices. Consequently, despite excess demand in some hours, consumers are unwilling to pay more, and the MCP remains below the cap, consistent with the law of demand.}

\begin{figure}[htbp]
\centering
\begin{subfigure}[t]{0.49\textwidth}
\centering
\includegraphics[width=\linewidth]{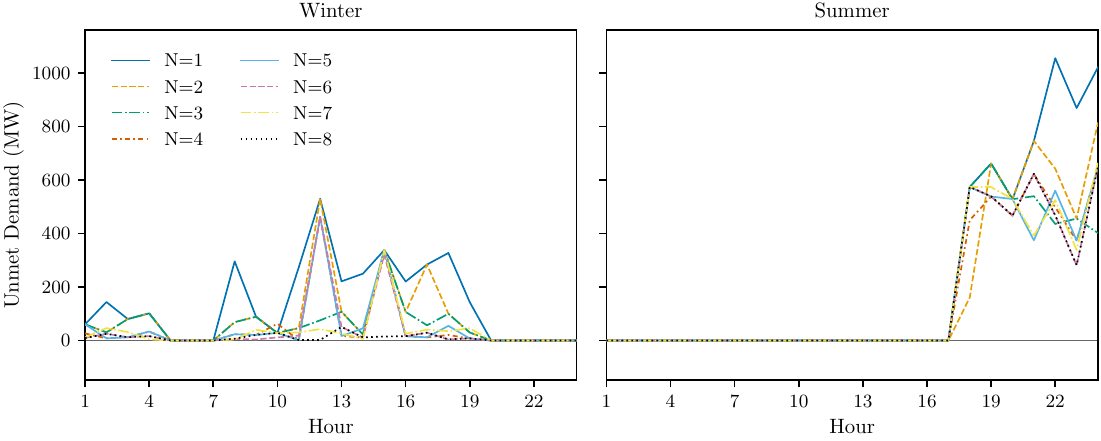}
\caption{Unmet demand under different levels of competition.}
\label{fig_unmet}
\end{subfigure}\hfill
\begin{subfigure}[t]{0.49\textwidth}
\centering
\includegraphics[width=\linewidth]{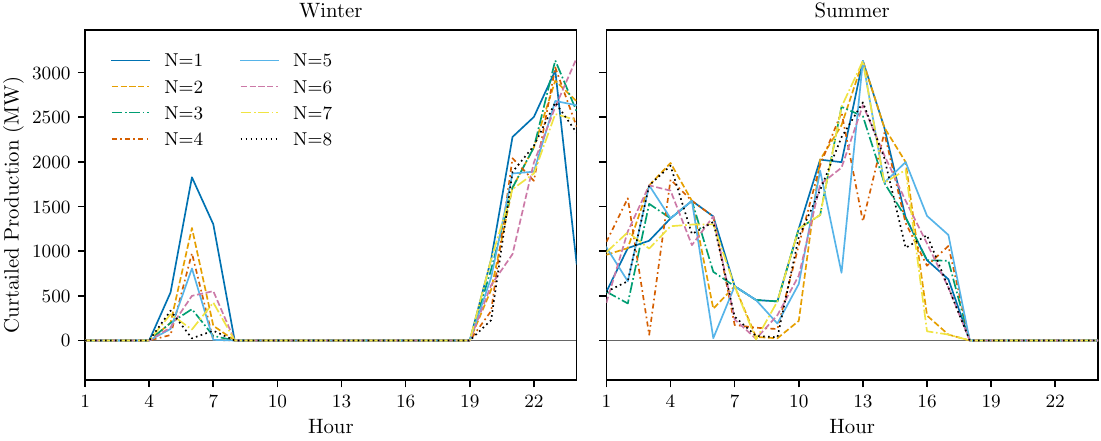}
\caption{Curtailed production under different levels of competition.}
\label{fig_curtail}
\end{subfigure}
\caption{Unmet and curtailed energy across competition levels: (a) unmet demand; (b) curtailed production.}
\label{fig_unmet-curtail}
\end{figure}

The results presented in Figure \ref{fig_unmet-curtail} clearly illustrate the system-wide impacts of increasing the number of storage operators from one to eight. Unmet demand (panel~a) is higher under the Monopoly case, particularly during evening hours in summer when demand exceeds both renewable supply and available discharge capacity. However, as the number of storage operators increases, unmet demand declines sharply, with the most significant improvements observed when there are four to six storage operators. Beyond this, adding storage players yields only marginal reductions, indicating diminishing returns in terms of reliability gains from additional market participants.

Conversely, Figure \ref{fig_unmet-curtail}(b) shows that curtailed production follows the opposite trajectory. With fewer operators, curtailment remains low because the actively utilized storage capacity is insufficient to absorb all renewable surpluses. As the number of operators rises and withholding decreases, total curtailment actually increases slightly in some scenarios, particularly during midday hours in summer when solar generation peaks and all storage units simultaneously hit their charging capacity limits. In winter, the effect on both unmet demand and curtailment remains relatively moderate throughout the day.

\subsubsection{Welfare and sensitivity analysis}
For the welfare and sensitivity analysis, we solve the social planner's problem under varying numbers of storage players $N$ and aggregate storage capability. The planner jointly optimizes renewable curtailment and storage dispatch over the 24-hour horizon, internalizing intertemporal constraints and removing strategic behavior. Changing $N$ reallocates a fixed total capability across $N$ units under coordinated dispatch, so it affects outcomes only through technical granularity. This establishes an efficiency frontier against which strategic storage operators can be evaluated, while scaling aggregate capacity identifies optimal sizing limits.

\begin{figure}[htbp]
\centering
\begin{subfigure}[t]{0.49\textwidth}
\centering
\includegraphics[width=\linewidth]{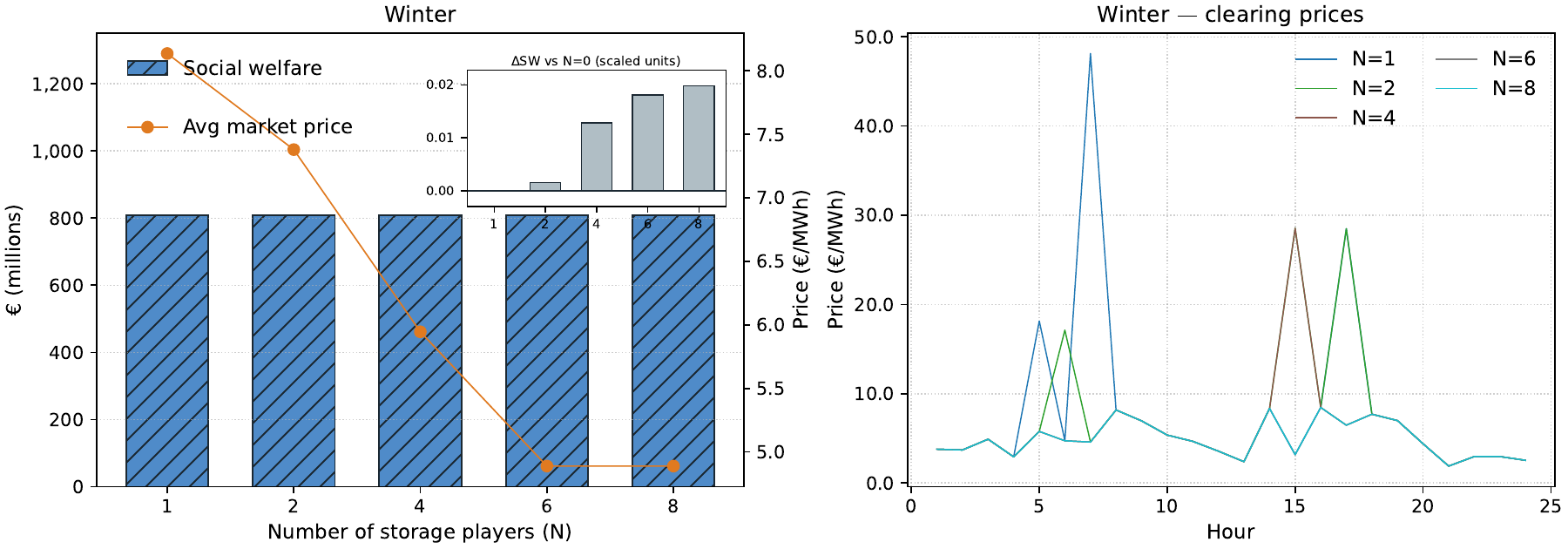}
\caption{Social welfare and electricity prices under different levels of competition: Winter.}
\label{SW_price_w}
\end{subfigure}\hfill
\begin{subfigure}[t]{0.49\textwidth}
\centering
\includegraphics[width=\linewidth]{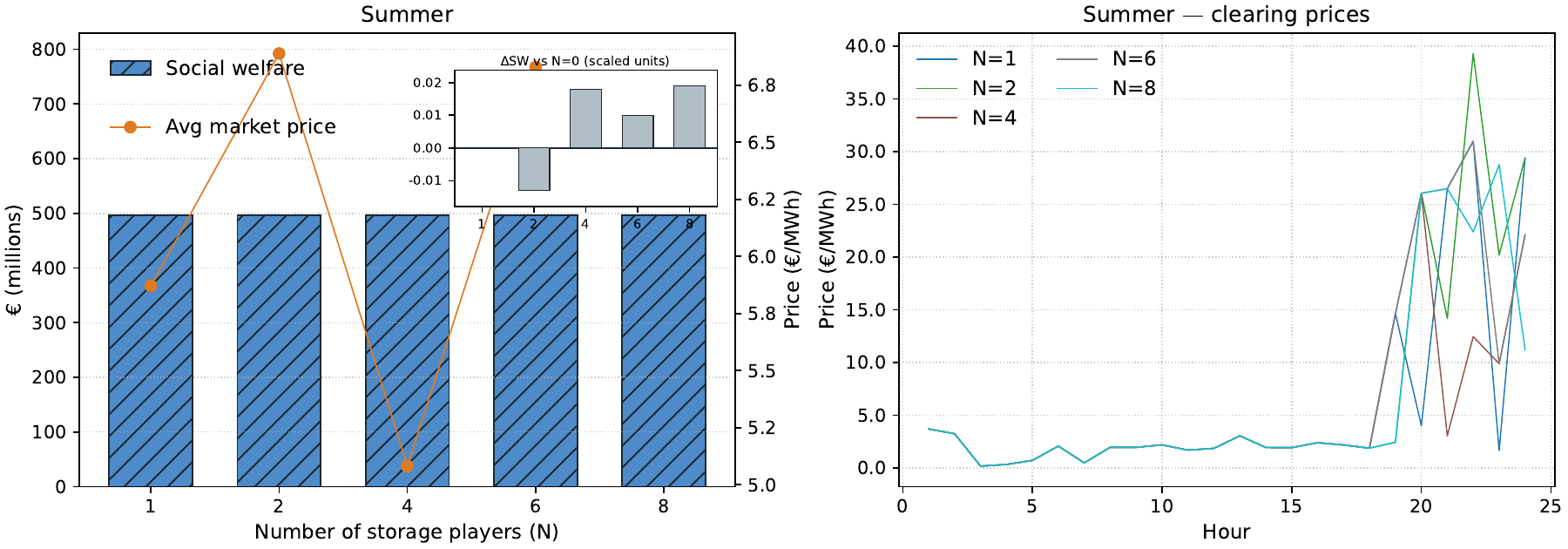}
\caption{Social welfare and electricity prices under different levels of competition: Summer.}
\label{SW_price_s}
\end{subfigure}
\caption{Social welfare and electricity prices by season under varying competition: (a) Winter; (b) Summer.}
\label{fig_SW_price_both}
\end{figure}

The results in Figure \ref{fig_SW_price_both} report total social welfare (summed over 24 hours). The line plots show the daily average hourly clearing prices $\{\lambda_t\}_{t=1}^{24}$. Because the effect of competition levels on total social welfare is relatively small compared to the overall welfare magnitude, an inset is provided to highlight the variation across the number of players, expressed in the same welfare units. 

The key takeaway is that partitioning a fixed aggregate storage capacity among a larger number of operators under a coordinated social planner does not inherently change total social welfare. 
From Figure \ref{SW_price_w}, it can be seen that: \textit{i)} total welfare is essentially flat with respect to $N$. The inset shows a small, asymptotic increase in social welfare, indicating that finer technical granularity improves welfare slightly, but under full coordination, the magnitude is modest. \textit{ii)} The average clearing price is generally lower with more subdivided capacity. Prices fall in high net-load hours and rise slightly in scarcity hours. Because winter net load is higher, the price effect is more visible than on a summer day, though the total social welfare change remains small. \textit{iii)} Most of the welfare gain accrues to consumer surplus through reduced peak prices, while changes in producer surplus are small after netting storage operating costs.

The summer results in Figure~\ref{SW_price_s} mirror the winter case: total welfare is nearly invariant to the number of capacity divisions, while average clearing prices are lower and less elastic than in winter, consistent with a smoother renewable profile. Overall, absolute welfare gains from partitioning are small in both seasons, whereas the price impact is meaningful in winter, where storage lowers peak prices when the net load is most volatile.

The social planner's adequacy outcomes, namely unmet demand and curtailment as a function of the number of storage divisions, together with the effect of scaling total storage capability under undersized, calibrated, and oversized scenarios, are reported in Sections~S6 and~S10 of the Supplementary material. Both reinforce the diminishing-returns pattern: a modest, well-coordinated amount of storage captures most of the system benefits, while further subdivision or oversizing yields little additional gain.

We also conduct sensitivity analyses on operating costs and battery efficiency in the social planner model, holding the number of storage operators fixed at $N=2$ and the capacity multiplier at $\theta=1.8$ (near the welfare-maximizing threshold). We vary battery efficiency as $\eta\in\{0.70,0.79,0.81,0.90,0.95\}$ and operating costs as $OC\in\{0.0,0.25,0.5,1.0,2.0\}$ (\EUR{}/MWh). In both winter and summer, total social welfare is largely insensitive over these ranges, and changes in $\eta$ have only negligible effects on average prices. By contrast, higher operating costs raise average prices modestly, by about 2–5 \EUR{}/MWh in summer and 5–7 \EUR{}/MWh in winter, and inevitably increase both unmet demand and renewable curtailment as marginal storage arbitrage becomes economically less attractive.

\subsection{Storage behavior under real-data scarcity (2024 case)}
\label{sec_scarcity-real}

In the baseline configuration, wholesale prices rarely reach extreme scarcity levels because the available supply (renewables plus storage discharge) typically intersects the first few high-volume steps of the inverse-demand curve. To complement this, we construct two real-data case studies using observed 2024 profiles (a summer day with a single storage operator and a winter day with two operators) that are explicitly selected to induce hours in which $\mathrm{RES}_h + Q^{\max}_{\mathrm{system}} < V_{\mathrm{cap},h}$, where $V_{\mathrm{cap},h}$ is the first cumulative-volume step at the administrative price cap. The full time-series results, hourly inverse-demand curves, and quantitative discussion are reported in Section~S8 of the Supplementary material. The two main findings from that analysis inform the interpretation of the 2030 results presented here. First, in hours where aggregate deliverability falls short of $V_{\mathrm{cap},h}$, the market clears at (or very close to) the administrative cap regardless of the ownership structure. Second, ownership primarily alters the \emph{allocation and timing} of charge/discharge rather than the occurrence of scarcity itself. This reinforces the structural distinction between \emph{energy capacity} (which enables intertemporal shifting) and \emph{power capacity} (which governs real-time adequacy), and motivates the policy observation that adequacy products or minimum-deliverability requirements targeted at critical hours are the appropriate instruments for curbing cap-price incidence in a 100\% RES market.

\section{Discussion}\label{sec_discussion}

This section summarizes the key results from the proposed model, and against existing literature to outline managerial and policy implications. \textit{i) Storage improves market efficiency even under strategic behavior.} Profit-maximizing storage arbitrage by charging during surplus periods and discharge during deficit periods, reducing production curtailment and unmet demand relative to a no-storage scenario. However, strategic operation does not fully flatten prices or close supply-demand gaps. In a fully renewable setting, some unmet peak demand and occasional trough curtailment persist, consistent with prior work showing that decentralized equilibria with strategic actors fall short of the social optimum \citep{Huang2022,nasrolahpour2016bidding}.

 \textit{ii) Market power induces capacity withholding and efficiency losses.} Oligopolistic storage operators strategically withhold capacity to support higher prices, resulting in a welfare loss compared to perfect competition or the social planner's benchmark. The results show that a single operator smooths fluctuations but leaves excess unserved demand and curtailment relative to competitive benchmarks \citep{8Coordinated,anunrojwong2024battery}. This price-making ability reflects the broader literature on strategic storage, where profit-maximizing batteries discharge below welfare-maximizing quantities and exploit system imbalances, even when underlying generation has near-zero marginal cost \citep{bjorndal2023energy,Schill2010}.
\textit{iii) Modest storage competition substantially diminishes} market power. The gap between Nash equilibrium outcomes and the social planner's dispatch shrinks rapidly as the number of independent storage operators rises. In our calibrated stylized model, with \emph{three} strategic operators, prices, welfare, and storage utilization become close to the social planner’s solution, and additional entry mainly redistributes surplus from storage owners to consumers. This aligns with oligopoly theory and recent analyses showing narrow price-of-anarchy bounds and sharp efficiency improvements from limited competition \citep{anunrojwong2024battery}. At the same time, the numerical cutoff should be interpreted as scenario-calibrated rather than universal.
 \textit{iv) Diminishing returns and capacity saturation}. More storage is not always better. In our calibrated case study, absolute welfare gains taper once total energy capacity reaches roughly \(1.0\text{--}1.5\times\) the residual demand, whereas doubling the capacity multiplier ($\theta$) is sufficient to virtually flatten arbitrage margins. Larger aggregate capacity also intensifies competition among storage units, eroding margins and potentially undermining investment incentives. These saturation effects resonate with prior findings \citep{johnson2024optimal,gaudard2019energy}. The implication is that socially optimal capacity sizing is critical to curb volatility and production curtailment, but it should not be so large that arbitrage opportunities are entirely exhausted and the business case collapses.

Overall, our simulation results and methodology isolate the strategic impact of storage operators. By analyzing a 100\% renewable context, we address a scenario that remains underexplored in prior strategic storage studies, which often include conventional generation or reserve markets (e.g., \cite{bjorndal2023energy,zhang2020cournot}). We isolate storage as the primary flexible, price-making resource and show that, even with zero-marginal-cost renewable generation, oligopolistic storage can induce nontrivial price distortions.

Methodologically, we compute Nash equilibria among multiple storage players in an MILP-based Cournot framework to endogenize strategic interactions more efficiently. Related multi-storage work includes capacity games and multi-node settings \citep{zhao2022strategic,Huang2022}. We quantify efficiency losses across competition levels and show they fall from material levels under a monopoly to $<\!1\%$ of total welfare under a triopoly, consistent with empirical validations of bounds in \cite{anunrojwong2024battery}.

\begin{figure}[H]
\centering
\includegraphics[width=0.95\textwidth]{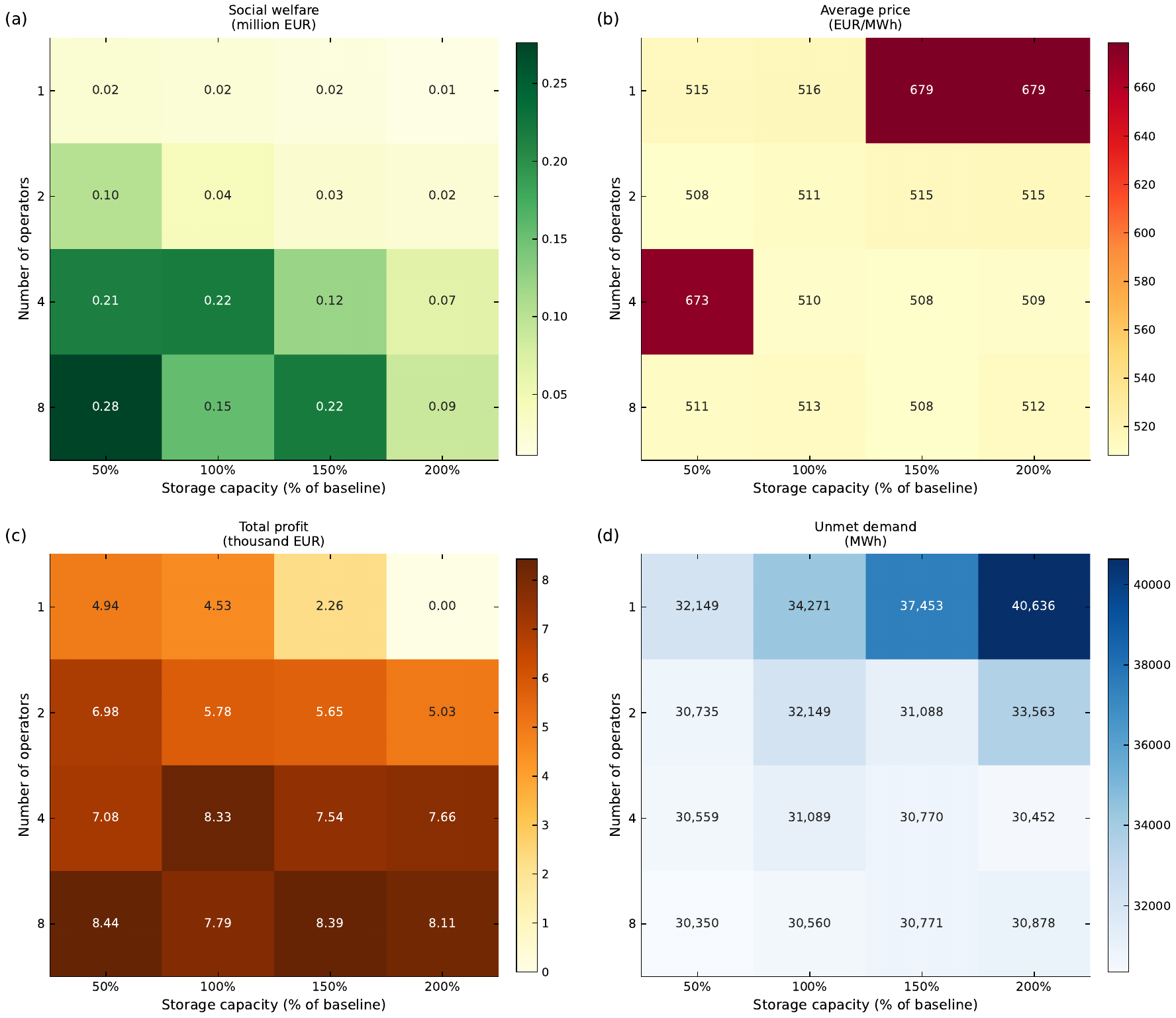}
\caption{Joint effect of storage competition ($N$) and aggregate capacity multiplier ($\theta$) on market outcome.}
\label{fig_management_heatmap}
\end{figure}

To translate the equilibrium analysis into actionable insights, we examine its implications for system planners and for private storage operators. Figure~\ref{fig_management_heatmap} maps the two primary market-design dimensions, the number of competing storage operators ($N$) and aggregate storage capacity ($\theta$), to social welfare, clearing prices, aggregate storage profits, and unmet demand. Two implications for system design follow. First, the marginal welfare gain from adding operators is largest when ownership is concentrated and declines as the number of operators grows. Beyond moderate competition, further additions have limited effect on clearing prices. Adding operators is therefore effective only in concentrated markets. Second, increases in aggregate capacity continue to reduce physical scarcity and compress price spreads across all competition levels. Aggregate capacity therefore becomes the decisive factor once competition reaches a moderate level. Each addresses a distinct mechanism. Adding operators disciplines strategic bidding. Increasing aggregate capacity relaxes physical deliverability. Competition cannot eliminate scarcity when aggregate capacity is insufficient. Capacity cannot discipline pricing under concentrated ownership. Configurations with both moderate competition and moderate aggregate capacity therefore dominate configurations that rely on either one alone. Three implications for storage operators follow from the same mapping. First, aggregate storage profits peak under concentrated ownership with moderate aggregate capacity. Profits decline as the market becomes either highly competitive or capacity-abundant. Operators should therefore evaluate expected profitability against the market configuration relevant to their own decision rather than against the baseline alone. Second, the equilibrium bidding behavior depends on market structure. In concentrated markets, individual bid curves meaningfully affect clearing prices and operators face a non-trivial strategic bidding problem. In competitive markets, the equilibrium converges toward price-taking and individual bids have little effect on clearing outcomes. Bidding sophistication is therefore consequential in concentrated markets but has limited effect on clearing outcomes in competitive markets. Third, in the two-player case, winter converges to a dominant equilibrium with one smaller secondary outcome, while summer admits a broader set of equilibria. The summer benchmark should therefore be treated as a point within a wider range of possible outcomes rather than as a single prediction. The qualitative relationships above reflect mechanisms general to high-RES electricity systems. Diminishing marginal welfare returns reflect the standard convergence of Cournot outcomes toward competitive outcomes as the number of operators grows. The partial substitutability of competition and aggregate capacity reflects the separation between pricing discipline and physical deliverability. The seasonal asymmetry reflects the concentration of residual-demand scarcity in winter. The numerical thresholds, however, are calibrated to the Denmark 2030 case study. The specific number of competitors required to discipline a given market, or the profitability of a given storage configuration, will depend on regional generation mixes, interconnection capacities, and local market-design rules.

Regarding investment, planners should recognize the inevitable phase of saturation. Overbuilding storage in energy-only markets compresses price spreads and jeopardizes capital cost recovery. Staged deployment and diversified revenue stacking (e.g., ancillary services, capacity markets) can sustain viability as energy arbitrage margins narrow. If society targets high storage penetration for reliability or decarbonization, but energy-market revenues become insufficient due to welfare-enhancing price smoothing, complementary mechanisms such as capacity remuneration, reliability options, contracts for difference \citep{abate2022contract}, or regulated returns may be warranted. The overarching goal must be to ensure efficient short-run dispatch (low volatility) while maintaining long-run investment incentives for the very assets that deliver that efficiency. Allowing limited scarcity pricing or providing separate capacity payments represents two coherent paths forward.

\textit{The proposed model has some limitations}: It is a deliberately stylized benchmark of storage behavior in a 100\%-RES, day-ahead, energy-only market. For tractability, we \textit{i}) ignore network details and locational pricing (no congestion or nodal effects), \textit{ii}) assume perfect foresight and a daily neutral state-of-charge return (no forecast risk or inter-day arbitrage), and \textit{iii}) fix the number, size, and risk preferences of non-degrading storage units (no endogenous entry/exit, heterogeneity). These choices mean that the quantitative results should be interpreted strictly within their defined scope. 

\section{Conclusion}\label{conclusion}

This study developed a game-theoretic model to investigate the strategic behavior of energy storage operators in a uniform-price day-ahead electricity market supplied 100\% by renewable energy sources (RES). A Cournot competition model was formulated to capture the profit-maximizing bidding strategies of storage operators, while a centralized social planner optimization provided the social-welfare benchmark. Two reformulations were proposed: one based on a big-$M$ method, and the other on a continuous demand-block reformulation. The model was applied to Denmark’s power system using 2024 renewable and aggregate demand data, scaled to 2030 renewable energy projections and DK1 bidding zone demand profiles, thereby combining realistic and forward-looking scenarios.

The results show that introducing strategic storage operators can significantly improve market efficiency by smoothing daily supply-demand imbalances. By charging during periods of high renewable output and discharging when renewable supply is scarce, storage arbitrages between low- and high-price hours. This reduces unmet demand, lowers renewable curtailment, and increases total welfare relative to the no-storage case, where prolonged surpluses would otherwise be curtailed and deficits would leave demand unmet. Accordingly, the dominant behavior is to charge in surplus hours and discharge in shortage hours. This pattern both maximizes operators’ profits and improves social welfare by minimizing curtailment costs and welfare losses from unmet demand.

 At the same time, strategic bidding can confer market power on storage operators and distort efficiency. Under monopoly or duopoly competition, operators can intentionally withhold capacity to wait for high-price periods, thereby raising their own profits while doing little to fully eliminate unmet demand or curtailment. This behavior predictably lowers welfare relative to the social planner benchmark under perfect coordination. As the number of competitors increases, these distortions diminish. In our market setting, the social planner benchmark is approached rapidly, with three competing operators already delivering outcomes close to the social optimum. Beyond that, additional entrants yield only marginal welfare gains, and the primary effect of competition becomes a surplus transfer from producers to consumers via lower prices. Therefore, this numerical threshold should be interpreted as scenario-calibrated rather than universal.

The analysis also underscores the importance of storage sizing. Undersized systems leave demand peaks and price spikes unmitigated. Beyond roughly $1.0$--$1.5\times$ the residual-demand benchmark in storage capacity, further expansion yields negligible system gains and primarily dilutes profits, weakening investment incentives and potentially harming long-term welfare. Thus, more storage is not always better. Moderate capacities capture the vast majority of benefits while preserving revenues and clear market signals. As with the competition result above, the precise saturation point depends on the scenario design and should be viewed as a benchmark magnitude rather than a universal rule.

Overall, the findings highlight the dual role of storage operators in high-RES markets. On the one hand, by smoothing renewable variability, storage is indispensable for improving system efficiency. On the other hand, without appropriate design, it can become a significant source of market power. Effective market design and regulation are therefore crucial for aligning private incentives with social welfare. The targeted robustness analyses further show that this central strategic mechanism is stable, even though the absolute levels of welfare, scarcity, and prices can shift with representative-day selection, demand-block granularity, and information assumptions. Managerially, the results suggest that ownership structure, and storage-capacity deployment should be evaluated jointly rather than separately. Tools such as concentration limits, must-offer obligations in scarcity hours, carefully designed scarcity-pricing rules, and complementary remuneration mechanisms can help preserve storage's operational value while curbing welfare losses from strategic withholding.

The paper can be extended in several directions. \textit{i)} Move beyond representative days to multi-day and seasonal horizons with richer terminal SoC specifications to capture intertemporal arbitrage, deeper cycling, and reserve interactions. \textit{ii)} Incorporate uncertainty in RES output and demand using stochastic or robust formulations, and test alternative demand reconstructions or supply-function variants for robustness. \textit{iii)} Endogenize investment and revenue sufficiency under alternative market designs, allowing for heterogeneous and risk-averse storage operation across locations in a nodal model with transmission constraints. 

% ==== Embedded bibliography (bibtex on references.bib). ====

\end{document}